\documentclass[article]{aa}
\usepackage{graphicx}
\usepackage{eufrak}
\usepackage{txfonts}

\begin{document}

\title{Mass Loss Evolution and the Formation of Detached Shells around TP-AGB Stars}
\titlerunning{Mass Loss Evolution and the Formation of Detached Shells around TP-AGB Stars}

\author{Lars Mattsson\inst{1}\thanks{\email{mattsson@astro.uu.se}}\and Susanne H\"ofner\inst{1}\thanks{\email{hoefner@astro.uu.se}}
        \and Falk Herwig\inst{2,3}\thanks{\email{fherwig@astro.keele.ac.uk}}}
\institute{Department of Astronomy and Space Physics, Uppsala University, Box 515, SE-751 20 Uppsala, Sweden
         \and Keele Astrophysics Group, School of Physical and Geographical Sciences, Keele University, Stafford shire ST5 5BG, UK \and Theoretical Astrophysics Group, LANL, Los Alamos, NM 87545, USA}

\offprints{Lars Mattsson}

\date{Received date; accepted date}

\abstract
{ The origin of the so called 'detached shells' around AGB stars is not fully understood,
  but two common hypotheses state that these shells form either through the interaction 
  of distinct wind phases or an eruptive mass loss associated with a He-shell flash. 
  We present a model of the formation of detached shells around thermal 
  pulse asymptotic giant branch (TP-AGB) stars, based on detailed modelling of mass loss and stellar evolution,
  leading to a combination of eruptive mass loss and wind interaction.}
{ The purpose of this paper is first of all to connect stellar evolution with wind and mass 
  loss evolution and demonstrate its consistency with observations, but also to show how thin detached
  shells around TP-AGB stars can be formed. Previous attempts to link mass loss evolution with the formation of detached 
  shells were based on approximate prescriptions for the mass loss and have not included detailed modelling
  of the wind formation as we do here. }
{ Using stellar parameters sampled from an evolutionary track for a $2 M_\odot$ star, 
  we have computed the time evolution of the atmospheric layers and wind acceleration 
  region during a typical thermal pulse with detailed radiation hydrodynamical models 
  including dust formation. Based on these results, we simulate the subsequent circumstellar envelope (CSE) evolution 
  using a spherical hydrodynamic model.}
{ We find that existing simple mass loss prescriptions all suggest different mass loss evolutions and that they differ 
  from our detailed wind modelling. The most important factor for the formation of a detached shell is 
  the wind velocity evolution which has a strong impact on the wind interaction and the 
  resulting pile-up of matter. Our CSE model shows that a thin shell structure may be formed 
  as a consequence of a rather short phase of intense mass loss in combination with a significant variation in the wind 
  velocity, as obtained by our wind models. This situation 
  can \it only \rm be obtained for a limited range of amplitudes for the piston boundary used in the dynamic atmosphere 
  models.}
{ The combined mass loss eruption and wind interaction scenario for the formation of detached shells around AGB stars 
  (suggested by previous work) is confirmed by the present modelling. Changes in mass loss rate and wind velocity due to 
  a He-shell flash are adequate for creating distinct wind phases and a 'snow plow effect' that is necessary to form a 
  geometrically thin detached shell. The derived properties of the shell (i.e. radius, thickness and density) are 
  more or less consistent with existing observational constraints. }

\maketitle

\section{Introduction}
  The origin of the so called 'detached shells' around TP-AGB stars in a late evolutionary stage is not fully understood. 
  As a phenomenon, these structures have been known for about 20 years (see e.g. \cite{Olofsson87}). The idea that they 
  were connected with He-shell flashes soon emerged, but the exact formation mechanism remained a matter of debate 
  (\cite{Olofsson90}). In fact, it is still not clear that detached shells are only of internal origin. Wareing et al. 
  (2006) \nocite{Wareing06} have recently shown that shell-like structures like these can emerge from the interaction of an 
  AGB wind with the interstellar medium (ISM) in which the star is moving. The shell structure could then be explained as 
  a bow shock into the ISM.
  
  A more common, older hypothesis for how these structures are formed, states that they arise from the interaction 
  between a fast and a slow wind (two-wind interaction) from the central star. As the fast wind propagates outwards, it 
  will sweep up material ejected through the slow wind and create a shell-like structure. This shell then contains matter 
  ejected by the fast wind, together with swept up matter ejected by the slow wind. In order to work, this hypothesis needs
  a sufficiently high mass loss rate and wind velocity associated with the fast wind, as well as a 
  non-negligable pre-flash wind (for more details, see \cite{Steffen98}, \cite{Olofsson00}, \cite{Schoier05} and references
  therein). Kwok et al. \nocite{Kwok78} (1978) presented  an analytic wind interaction model in which
  a fast wind is running into a slower wind -- each having a constant mass-loss rate and constant velocity. Hence, 
  the distance from the central star to the interaction region is expected to increase linearly over time. This simple 
  picture has been used to link observed mass loss properties with the presence of detached shells 
  (\cite{Olofsson93}, \cite{Schoier01}, \cite{Schoier05}). 

  Another early hypothesis (\cite{Zuckerman93}) for the formation of thin shell structures suggests
  that a short event of high mass loss (with a non-evolving wind velocity), which we refer to as a "mass loss eruption", 
  would simply translate into a correspondingly narrow, high-density shell moving through the circumstellar envelope. 
  This, however, was shown to be an insufficient mechanism for the formation of detached shells (\cite{Steffen97}, 
  \cite{Steffen98}). But the same mass loss evolution could still produce the detached shell phenomenon if the wind before 
  the flash is slower than the wind during the eruption, due to the sweep-up of matter that occurs as the matter released by 
  the mass loss eruption propagates through the CSE created by the previous mass loss. This mass loss eruption scenario is 
  most likely tied to a He-shell flash (thermal pulse) as suggested by Olofsson et al. \nocite{Olofsson90} 
  (1990) and Vassiliadis \& Wood \nocite{Vassiliadis93} (1993). Under the assumption that the slow 
  (pre-flash) stellar wind represents an ordinary state of mass loss for a TP-AGB star there is still 
  the need of a period of relatively high mass loss and an enhanced outflow velocity, also in an 
  interacting wind scenario. 

  Previous numerical models of the connection between a He-shell flash and the formation of a
  detached shell have provided some support for the wind-interaction hypothesis, but as mentioned
  above, only as a partial explanation. Steffen \& 
  Sch\"onberner \nocite{Steffen00}(2000), from hereon abbrevated SS2000, confirmed this picture of
  a detached shell forming from a mass loss eruption and subsequent wind interaction qualitatively, 
  using a spherical hydrodynamic model, altough the mass loss evolution is prescribed, not modelled,
  i.e. a more or less \it ad hoc \rm assumption about the inner boundary. 

  A rather detailed 
  3D model of the hydrodynamical features of a detached shell was presented by Myasnikov et al. 
  \nocite{Myasnikov00}(2000), where the gasdynamic features of detached shells around a carbon rich 
  TP-AGB star with variable mass loss rate were investigated numerically. They find that these shells 
  are unstable to turbulence and that Rayleigh-Taylor instabilities develop in the radiative case. 
  Furthermore they find that 2D calculations show quite different features compared with the 3D case
  and that the structures of the predicted shells are dependent on the initial basic wind properties. 
  The latter is a very important finding. In order to confirm the two-wind interaction hypothesis 
  on a more detailed level, it is thus cruical to connect the evolution of the stellar parameters 
  $(M,L_\star,T_{\rm eff})$ with the mass loss and wind evolution during the He-flash event.
  
  To understand the mass loss evolution of TP-AGB stars it is important to know how stellar
  evolution affects the physics of the atmosphere and consequently the formation of a stellar
  wind. This can be done by feeding the stellar parameters resulting from a stellar evolution
  model into a dynamic stellar atmosphere model. The models for the atmospheres and winds used in the 
  present paper are computed using frequency-dependent radiative transfer for the gas and dust, including 
  detailed micro-physics of the dust 
  grains and their formation (described in \cite{Hofner03} and \cite{Andersen03}) in combination
  with time-dependent hydrodynamics. It has been shown that synthetic opacity sampling spectra for 
  carbon-rich AGB stars based on these models reproduce the observed spectral energy
  distributions of these stars quite well (\cite{Gautschy04}). Furthermore, for the first time, these models
  are also able to simultaneously reproduce the observed time-dependent behaviour of fundamental, first {\it and}
  second overtone rotation-vibration lines of CO, which are formed in the outflow, wind acceleration region 
  and atmosphere respectively (Nowotny et al. 2005a, 2005b). The mass loss properties, however,
  are more difficult to compare with observations. Derivations of 'observed' rates of mass 
  loss (e.g. \cite{Schoier05}) rely on stellar parameters and absolute
  distances that can be uncertain in many cases, which means that observations can only provide 
  limited constraints on mass loss rates and dust-to-gas ratios. Modelling detached shells might 
  therefore be an independent test of the mass loss properties of our dynamic atmosphere model.
  
  In this paper we combine this radiation-hydrodynamics (RHD) model of a pulsating TP-AGB atmosphere and wind 
  acceleration region with a detailed stellar evolution model and a simple spherical hydrodynamic model to follow
  the dynamical evolution of the CSE on long time scales and study the formation of a geometrically
  thin detached shell around the star. We are thus dealing with three different scales: the internal evolution, the
  atmosphere and the effects on the CSE. As this paper shows, our theoretically expected wind properties 
  are such that a thin detached shell is formed as a natural consequence of a He-shell flash.

\section{Theory and Method}
\subsection{General Description}
  \label{general}
  On a descriptive level, the basic idea behind the modelling is to connect models of three different regimes 
  (see Table \ref{scheme}). 
  The starting point is a pre-computed model of the internal evolution of a star with a main sequence mass of 
  $2M_\odot$ and an initial metallicity of $Z = 0.01$. This track is not chosen for any particular reason, other
  than that it is consistent with the type of carbon stars with known detached shells.
  The resulting stellar parameters $(M,L_\star,T_{\rm eff}, {\rm C/O})$ are sampled with suitable time intervals around 
  a thermal pulse. These parameter values are then fed into an RHD atmosphere model that provides us with 
  the mass loss properties. Knowing these properties we can construct a time dependent inner boundary condition 
  for a hydrodynamic model of the larger-scale circumstellar evolution, i.e. the formation of a detached shell 
  from the interaction of different wind phases. The reason why we split up the modelling in this way, is the 
  very different time and length scales of the relevant physical processes involved. An 'all-in-one' model would 
  demand unreasonable computing power in order to resolve the smallest lenght scales and the shortest time scales. 
  In this section we describe the models and how the computations are carried out.

\subsection{Dynamic Atmospheres}
  \label{dynamic}
  The 1D (spherically symmetric) RHD code predicts the mass-loss rates of TP-AGB stars by detailed modelling of the 
  atmosphere and the circumstellar environment around these pulsating long-period variable stars. The model 
  includes frequency-dependent radiative transfer and dust formation, i.e. we are solving the coupled system of 
  frequency-dependent radiation hydrodynamics and time-dependent dust formation (cf. \cite{Hofner03}) employing 
  an implicit numerical method and an adaptive grid.

  In our model the stellar atmosphere and circumstellar envelope are described in terms of conservation 
  laws for the gas, the dust and the radiation field, i.e. we solve the following set of coupled, nonlinear partial 
  differential equations (PDEs):

  \begin{itemize}
  \item The three equations describing conservation of mass, momentum and energy for the gas.
  \item The 0th and 1st moment equation of radiative transfer.
  \item The four moment equations of dust formation (cf. Section \ref{dustform}).
  \item The Poisson equation (self gravity).
  \end{itemize}

  To this system of nonlinear partial differential equations we add a ``grid equation'' which determines the 
  locations of the grid points according to accuracy considerations (\cite{Dorfi87}) and an equation keeping 
  track of the condensable amount of carbon, leaving us with a total of 12 partial differential equations (PDEs) 
  to solve. This system of PDEs is then solved implicitly using a Newton-Raphson scheme. All equations are 
  discretised in a volume-integrated conservation form on a staggered mesh. The spatial discretisation of the 
  advection term is a monotonic second-order advection 
  scheme (\cite{vanLeer77}). The same order of numerical precision is used for all PDEs. Details of the numerical 
  method are discussed by Dorfi \& Feuchtinger \nocite{Dorfi95} (1995) and in several previous papers about 
  dust-driven wind models (cf. \cite{Hofner95} and references therein).

  The dynamical calculations are carried out as follows. All wind models are started from hydrostatic dust-free 
  initial models where the outer boundary is located close to the photosphere (about $1.3 - 1.6 \,R_\star$). In 
  these initial models, the fundamental stellar parameters $(M,L_\star,T_{\rm eff}, {\rm C/O})$ can be chosen freely as 
  inital conditions for the models. In the present case we pick the sets of stellar parameters from an evolutionary 
  track as described in Section \ref{general}. When the dust equations are switched on, dust formation starts and  
  creates an outward motion of the dust and the gas. In the first computational phase the expansion of the atmospheric 
  layers is followed by the grid to about $20 - 30\,R_\star$ (usually $\sim 10^{15}$ cm). At this radius the location
  of the outer boundary $R_{\rm out}$ is fixed allowing outflow. The outflow model then evolves for typically more than 
  200 years, but the precise time is dependent on the pulsation period. The amount of mass between the inner boundary
  $R_{\rm in}$ and the outer boundary $R_{\rm out}$ is roughly $10^{-2}M_\odot$, which is sufficient to avoid any
  significant mass depletion within the time series considered here.

  \begin{table*}
  
  \caption{\label{scheme} Schematic overview of the model. Radial and time scales given are approximate figures.}
  \center
  \begin{tabular}{lccccl}
  \hline
  \\
  Model & Input & Output & Radial scale & Time scale & Comment\\
  \\
  \hline
  \\

  Stellar evolution & $M_{\rm ZAMS}$, $Z$ & $M_\star$,$L_\star$,$T_{\rm eff}$, ${\rm C/O}$ 
			  & $R\leq R_\star$ & $\sim 10^4$ yrs  
                    & Pre-computed \\[3mm]
  Atmosphere / wind & $M_\star$,$L_\star$,$T_{\rm eff},$ & $\dot{M}$, $v_{\rm wind}(r)$ 
                    & $R_\star<R \le 25 R_\star$ & $\sim 10^2$ yrs & Full RHD + dust, \\
                    & ${\rm C/O}$,$P$,$\Delta v_{\rm p}$ & & & & "snap shots" of evolution\\[3mm]
  CSE evolution     & $\dot{M}$, $v_{\rm wind}$ & $\rho_{\rm env}(r)$, $v_{\rm env}(r)$ 
                    & $25 R_\star < R_{\rm s} \le 10^4 R_\star$ 
                    & $\sim 10^4$ yrs & Gas dynamics,\\
                    & & & & & dust dynamics not considered\\

  \\
  \hline
  \\
  \end{tabular}
  \label{models}
  \end{table*}

  The model is self-consistent as far as the physics of the stellar atmosphere is concerned, but it does not include a 
  physical model for the pulsation mechanism. The pulsation is modelled phenomenologically using a 'piston boundary 
  condition' (\cite{Bowen88}) located at $R_{\rm in} \approx 0.9 R_\star$. To restrict the number of free parameters, 
  we employ an empirical period-luminosity relation (\cite{Feast89}) and keep the period locked to the luminosity for all 
  models. The piston amplitude, however, is essentially a free parameter that we cannot do away with as easily. For this 
  reason we must consider the effects of picking different amplitudes, which is a complication that we will get back to in 
  Section \ref{mlrdis}. The motion of the piston is accompanied by luminosity variations, since the flux through the
  lower boundary is kept fixed (as described in previous papers, e.g., \cite{Hofner97}). Note, however, that assuming no 
  artificial scaling of the short-term luminosity variations due to the pulsations, these are more than a factor of two 
  weaker than the long-term luminosity variation due to the flash.

\subsection{Dust Formation and Mass Loss}
  \label{dustform}
  Mass loss from TP-AGB carbon stars is mainly due to momentum transfer from radiation to atmospheric dust grains. This
  is often referred to as a {\it dust driven wind}. As such, a wind model for TP-AGB stars is highly dependent
  on how dust formation is treated. Hence, we will give a somewhat more detailed description of the dust model.  

  The models presented here include a time-dependent description of dust grain growth and evaporation using 
  the moment method by \nocite{Gail88} Gail \& Sedlmayr (1988) and \nocite{Gauger90} Gauger et al. 
  (1990). The dust component is described in terms of moments $K_j$ of the grain size distribution function 
  weighted with a power $j$ of the grain radius. The zeroth moment, $K_0$, is the total number density 
  of grains (simply the integral of the size distribution function over all grain sizes), while $K_3$ is 
  proportional to the average volume of the grains. The equations, which determine the evolution of the dust 
  component, are
  \begin{equation}
  \frac{\partial K_0}{\partial t} + \nabla\cdot(K_0\,v) = \mathfrak{J}, 
  \end{equation}
  \begin{equation}
  \frac{\partial K_j}{\partial t} + \nabla\cdot(K_j\,v) = \frac{j}{3} \frac{1}{\tau}K_{j-1} + N_\ell^{j/3}\mathfrak{J}, 
  \quad 1\le j \le 3,
  \end{equation}
  where $v$ is the flow velocity of the dust component, $\mathfrak{J}$ is the net grain formation rate per volume, $1/\tau$ 
  is the net growth-rate of the dust grains, and $N_\ell$ is the lower size limit of grains contributing to $K_j$.

  In order to calculate how much of the radiative energy and momentum is transferred to dust grains, we need to know the 
  frequency-dependent opacity of these grains. This can be expressed in terms of the extinction efficiency $Q_{\rm ext}$, 
  which is the ratio of the extinction cross section to the geometrical cross section of the grains. Since the grains are 
  usually small compared to the wavelengths,  $Q_{\rm ext}$ becomes a simple function of the grain radius. The wavelength  
  and grain size dependence of the opacity can thus be separated into two independent factors, which greatly simplifies the 
  calculations. The models in this paper are calculated using the data of \nocite{Rouleau91} Rouleau \& Martin (1991) 
  for $Q_{\rm ext}$ (see \cite{Andersen03} and \cite{Hofner03} for further discussion). The intrinsic dust density used 
  in the model is set to $\rho_{\rm d} = 1.85$ g cm$^{-1}$, which matches the material in Rouleau \& Martin (1991).

  We assume spherical dust grains consisting of amorphous carbon only, since the star we are modelling is a carbon star
  (see \cite{Andersen03} for details about these assumptions). The nucleation, growth and evaporation of grains is 
  assumed to proceed by reactions involving ${\rm C}$, ${\rm C_2}$, ${\rm C_2H}$ and ${\rm C_2H_2}$. 
  In this model of grain growth, a so called \it sticking coefficient \rm (sometimes referred to as the reaction efficiency 
  factor) is used, that enters into the net growth rate of the dust grains. This parameter, $\alpha_{\rm S}$, is not 
  definitely known unless we know the exact sequence of chemical reactions responsible for the dust formation. However, 
  Gail \& Sedlmayr (1988) argued that the sticking coefficient must be on the order of unity, mainly because it is expected
  that neutral radical reactions play a major role in the formation of carbon grains. It should be pointed out as well, 
  that with $\alpha_{\rm S} = 1$, our models nicely reproduce the expected mass loss properties of TP-AGB stars 
  (\cite{Schoier01}, \cite{Schoier05}). See Section \ref{mlrdis} for some further details. 

  Dust grains in a stellar atmosphere influence both its energy and momentum balance. For simplicity we assume complete 
  momentum and position coupling of gas and dust, i.e. the momentum gained by the dust from the radiation field is 
  directly transferred to the gas and there is no drift between dust and gas. However, this strong coupling between 
  the dust and the gas phase is not obvious. In a previous attempt to relax this phase coupling approximation, 
  \nocite{Sandin03, Sandin04} Sandin \& H\"ofner (2003, 2004) found that the effects of 
  decoupling of the phases might be quite significant. The most striking feature is that the dust formation may
  increase significantly, but this does not seem to necessarily increase the predicted mass loss rates for a given set of 
  stellar parameters. However, these models are not directly comparable with observations, since one essential 
  component -- frequency dependent radiative transfer -- is not included. Comparison with observations (\cite{Gautschy04}, 
  \cite{Nowotny05a}, \cite{Nowotny05b}) seems to indicate that in general the inclusion of frequency-dependent 
  radiative transfer is crucial for a realistic description of the pulsating amosphere, and therefore for the conditions in
  the wind formation region. A coupled solution of the detailed frequency-dependent equations of RHD, including dust 
  formation and drift would lead to a prohibitive computational effort and is not a realistic option in the present context.
  Thus, we chose to give frequency-dependent radiative transfer priority over relaxing the phase coupling approximation, 
  even if this approximation cannot hold for very low gas densities, i.e. at large distances from the star or for models 
  with low mass loss rates. Furthermore, since transfer of internal energy between gas and dust is negligible 
  compared to the interaction of each component with the radiative field (\cite{Gauger90}), we assume radiative equilibrium
  for the dust. This allows us to estimate the grain temperatures from the radiation temperature as we know the dust 
  opacities.

  From the models described in this section and in \ref{dynamic} we obtain the density and the wind velocity, both as 
  functions of radius and time. The mass loss rate is then given by
  \begin{equation}
  \label{massloss}
  \dot{M} = 4\pi R_0^2\,\rho_0(t)\,v_0(t),
  \end{equation}
  where $\rho_0$ and $v_0$ are the density and wind velocity at $R_0$, i.e. the radius where the outer boundary is 
  fixed, which is usually around $10^{15}$ cm and enough to ensure that the wind has reached the terminal velocity. 
  Since practically all momentum transfer from radiation to matter is due to the interaction with dust, we expect the 
  wind velocity to be correlated with the degree of dust condensation. Thus, Eqn. (\ref{massloss}) leads to the 
  expecation that under most circumstances a larger proportion of dust will correspond to a higher mass loss rate.

\subsection{Stellar Evolution}

\begin{figure*}
  \includegraphics[width=17cm]{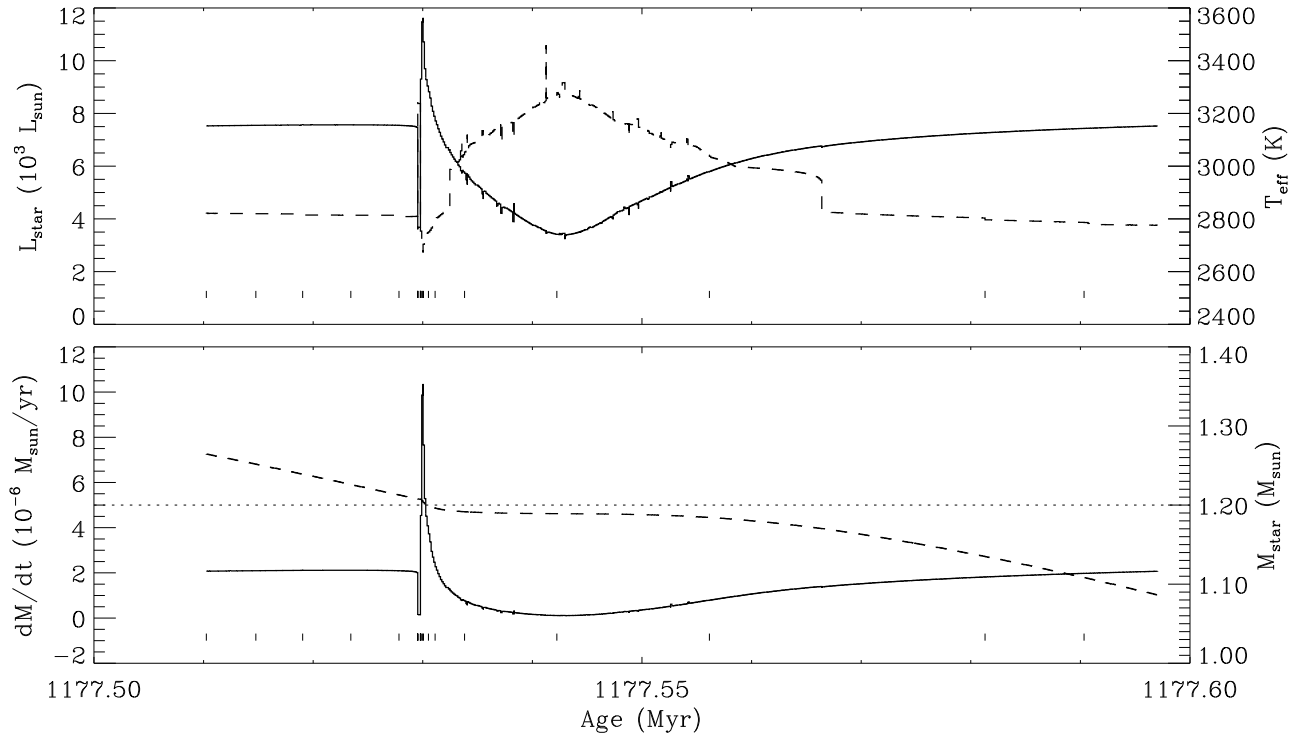}
  \caption{Time evolution of luminosity $L_\star$ (upper panel, full line), effective temperature $T_{\rm eff}$ (upper panel,
  dashed line), mass loss rate $dM/dt$ assumed in the evolution model (lower panel, full line)  and stellar mass $M_\star$ 
  (lower panel dashed line). The sampling points for the stellar parameters used as input to the dynamical atmosphere model 
  are tick-marked in the lower part of each panel and the dotted line in the lower panel indicates the stellar mass that
  we adopt for the atmosphere models. The jump in $T_{\rm eff}$ seen near $t=1177.58$ Myr is a numerical artifact.
  \label{tracks}}
  \end{figure*}

  To model how the atmosphere evolves on longer time scales (significantly longer than the pulsation
  period) due to changes in the internal constitution and nuclear processes of the star, we have 
  pre-computed a stellar evolutionary track (see Fig. \ref{tracks}). The stellar parameters are then 
  sampled at critical or characteristic points (tick-marked in the lower part of each panel in Fig. 
  \ref{tracks}) around a thermal pulse in the evolutionary track and used as input to the atmosphere 
  model. In total we pick 21 points along the track that we know from experience will capture the important features.
  By computing a set of models in this way for different points in the time evolution, we can 
  obtain the change of atmospheric properties and wind characteristics (i.e. mass loss rate and wind velocity) during 
  the flash event.

  The calculations have been done with the one-dimensional stellar
  evolution code EVOL (Herwig 2000, 2004 and references therein). We
  use the EVOL implementation of exponential, time and depth dependent
  overshooting, here with an efficiency $f = 0.016$ at the bottom of
  the convective envelope, but no overshooting at the bottom of the
  He-shell flash convection zone. EVOL features updated micro-physics
  input, including a nuclear reaction network. The Schwarzschild
  criterion for convection is used and the mixing length parameter is
  set to $\alpha_{\rm MLT} = 1.7$. Mass loss during AGB evolution is
  included as prescribed by \cite{Blocker95}. Specifically we use Eqn.\ (11) from that work
    in which $\Gamma(M)$ is determined according to his Eqn.\ (13) and
    (14). For the mass in Eqn.\ (13) the intial ZAMS mass is assumed as an
    approximation of the actual mass at which the oscillation period
    is $P_0 = 100\mathrm{d}$. $\dot{M}_R$ in Eqn.\ (11) is the Reimers
    mass loss rate (\cite{Reimers75}) as given by Eqn.\ (1) in
    \cite{Blocker95}. The scaling factor in that equation is set to
    $\eta_{\rm R} = 0.1$, and gives an overall evolution which is
    consistent with many obserational contraints, e.g., the number of
    thermal pulses expected after transformation into a carbon star.

  For this detailed study we have picked the last but one (the 17th out of 18) thermal pulse of sequence ET13 
  (\cite{Herwig06}). This sequence has the metalicity $Z=0.01$ and an initial mass of $2M_\odot$. At the time of the 
  selected thermal pulse the total mass is $1.22M_\odot$ and the core mass is $0.59M_\odot$. The third dredge-up is well 
  developed after the selected thermal pulse, with an efficiency of $\lambda = 0.154$, after $\lambda = 0.372$ at the 
  previous pulse. Dredge-up efficiency decreases at the end of the TP-AGB evolution when the envelope mass decreases.

  To minimise the number of atmosphere models needed to obtain the time evolution, we assume a constant stellar mass 
  ($M_\star = 1.2 M_\odot$) during the thermal pulse. This is a reasonable simplification in this context even if the 
  stellar mass will decrease notably on longer time scales due to the steady mass loss before the first flash and in 
  between subsequent flashes (see Fig. \ref{tracks}, which captures the flash event we have modelled). This assumption will
  enable us to use some models at more than one sampling point, since in contrast to, e.g., $T_{\rm eff}$, the stellar mass
  $M_\star$ is not a very critical parameter in the wind and CSE models. For the luminosity and temperature this 'recycling'
  of models leads to deviations from the evolutionary track no larger than $1\%$. Exact matching of stellar parameters will
  not increase the accuracy of the output, since small changes ($<1\%$) in the input parameters have virtually no effect on
  the resulting wind and mass loss rate. We also assume a constant ${\rm C/O}$ ratio (where 
  $\log(\varepsilon_{\rm O})$ = -3.35) before and after the flash event, where ${\rm C/O}=1.85$ before the flash, and 
  ${\rm C/O}=2.01$ during and after the flash. In practice, this is really not an approximation, since on an evolutionary 
  time scale the carbon abundance in the atmosphere makes a sudden jump from one basically constant level to another when a
  He-shell flash takes place. The numbers given above correspond to the evolutionary track we have computed.

\subsection{CSE Evolution}
  \label{cse_evol}
  Since the mass loss rate and wind velocity are changing with time, so are the boundary conditions for the modelling of 
  the CSE around the star (the region where the detached shell is formed). The inner boundary conditions for such a model, 
  i.e. wind velocity $v_0$ and density $\rho_0$ at $R_0$ (the radius where the mass loss is obtained) as functions of time,
  are defined by interpolating the time sequence of atmosphere models computed as described above. Here, we 
  are interested in whether these time variations will lead to the formation of a detached shell. The next step is 
  therefore to compute the dynamical CSE evolution subject to such variable boundary conditions. 

  Some combinations of stellar parameters and piston amplitudes $\Delta v_{\rm p}$ will result in models without any mass 
  loss (see Table \ref{parameters}). For such cases -- if they occur in the post-flash phase -- we assume a minimal mass 
  loss rate of $10^{-8}M_\odot$ yr$^{-1}$, which is motivated by observations (\cite{Olofsson98}, \cite{Schoier01}, 
  \cite{Schoier05}), and estimate the mimimum wind velocity to be $\sim 1$ km s$^{-1}$. The circumstellar density at $R_0$ 
  can thus be estimated using Eqn. (\ref{massloss}). Another possibility would be to simply take the average density at 
  $R_0$ from the dynamical model with the the outer boundary radius $R_0$ set closer ($\sim 1.5 R_\star$) to the 
  photosphere and assume a minimal mass loss rate, say $10^{-8}M_\odot$ yr$^{-1}$. Then the \it wind velocity \rm can be 
  estimated through Eq. (\ref{massloss}). We prefer the first method, where we set $v_0 = 1$km s$^{-1}$, although both 
  methods give more or less the same numbers. In the transition region between intense dust-driven mass loss and no 
  mass loss, we set the rate to $10^{-7}M_\odot$ yr$^{-1}$ and assume the wind velocity to be $3$ km s$^{-1}$ and compute 
  the density as described above. The exact numbers are not important in this case, since these models represent a phase 
  with essentially no stellar wind that occurs after the He-shell flash.  

  We employ a simplified model of the CSE evolution where the equations of continuity and motion are given by
  \begin{equation}
  \frac{\partial \rho}{\partial t} + \nabla\cdot(\rho v)  =  0
  \end{equation}
  \begin{equation}
  \frac{\partial \rho v}{\partial t} + \nabla\cdot(\rho v\,v)  =
  -\nabla P_\mathrm{gas},
  \end{equation}
  where $\rho$ is the mass density, $v$ is the velocity and $P_\mathrm{gas}$ is the thermal gas pressure. These 
  equations are solved using a simplified setup of the dynamic atmosphere code (radiative transfer and dust equations 
  switched off), i.e., the numerical method is the same. We assume that the gravitational pull from the star does 
  not affect the expansion of the CSE on the typical distance from the star considered here ($R\ge 10^{15}$ cm), since at 
  this radius it is orders of magnitude weaker than in the atmospheric layers. The self-gravity of the CSE can also be 
  considered negligible as the typical mass density is too low to produce any sigificant self-gravity effects. We also
  assume that the radiation field is dominated by the central star, i.e. a point-source field falling off as $\sim r^{-2}$.
  Hence, at large distances from the central star, neither gravity, nor radiation, will be a force that really matters. 
  This is of course a somewhat simplistic approach, but it is certainly a fair approximation of the physical situation in 
  the present case. The pressure gradient, however, is kept in the equation of motion, since it may play an important
  role in shock waves occuring at the interfaces of different wind phases.

  Regarding the temperature structure, we consider two limit cases.
  \begin{enumerate}
  \item [I:] No radiative heating or cooling, the temperature structure is set by the expansion of the gas,
  leading to a fast decrease of temperature with distance from the star. Shocks are assumed to be adiabatic.
  \item [II:] A fixed, given temperature structure only reflecting the radiation field of the star,
  leading to a more gentle dropping of temperature with distance.
  Isothermal shocks (relative to the surrounding medium).
  \end{enumerate}

  More precisely, in Case I we are solving an energy equation of the form
  \begin{equation}
  \label{energy}
  \frac{\partial \rho e}{\partial t} + \nabla\cdot(\rho e\,v)  =
  -P_\mathrm{gas} \nabla\cdot v ,
  \end{equation}
  (where $e$ is the specific internal energy and all other variables are denoted as before)
  together with the equation of motion and the continuity equation (see above).
  This means that the energy equation used for the CSE differs from the full RHD
  version of the atmosphere and wind acceleration region by the lacking source and sink
  terms for radiative heating and cooling (cf., e.g., H\"ofner et al.2003).

  In Case II, representing the other extreme, we assume that the
  temperature structure is fully determined by radiative equilibrium
  with the stellar radiation field. In this case, we have
  \begin{equation}
  \int_0^\infty \kappa_\nu(J_\nu - S_\nu)\,d\nu = 0,
  \end{equation}
  where $\kappa_\nu$ is the monochromatic opacity, $J_\nu$ is the mean intensity and $S_\nu$ is the source function.
  Hence,  the zeroth order moment equation of radiation,
  \begin{equation}
  \nabla\cdot H + \rho(\kappa_{\rm J}J_\nu - \kappa_{\rm S}S) = 0,
  \end{equation}
  where the frequency integrated opacities $\kappa_J$ and $\kappa_S$ are defined as in H\"ofner et al. (2003),
  implies that $H$ in such a case must be divergence free, e.g. like a point source field. 
  We thus estimate the radiative field from
  \begin{equation}
  J_\nu \approx W(r) B_\nu(T_\star),
  \end{equation}
  where
  \begin{displaymath}
  W(r)=\frac{1}{2}\left[1 - \sqrt{1 - \left(\frac{R_\star}{r}\right)^2}  \right]
  \approx \frac{1}{4}\left(\frac{R_\star}{r}\right)^2,\quad r\gg R_\star,
  \end{displaymath}
  is a geometrical dilution factor (cf. \cite{Lamers99}). If we assume radiative equilibrium for dust, and $T_{gas}\approx T_{dust}$, 
  we end up with the relation
  \begin{equation}
  \label{radeq}
  T_{\rm gas}^4\, Q_{\rm Pl}(a,T_{\rm dust}) = W(r)\,T_\star^4\,Q_{\rm Pl}(a,T_\star),
  \end{equation}
  where $Q_{\rm Pl}(a,T)$ is the Planck mean extinction efficiency as a function of the grain radius $a$ and the temperature.
  A simple estimate of the frequency dependence\footnote{If the monochromatic $Q$ is frequency-independent (gray case), 
  the Planck mean $Q_{\rm Pl}$ is independent of $T$ and solving Eqn. (\ref{radeq}) for $T_{\rm dust}$ gives
  $T_{\rm dust}\sim T_\star\,r^{-1/2}$ for $r \gg R_\star$. Opacities are, however, frequency dependent and $Q_{\rm Pl}(a,T)$ 
  is therefore not a constant.} can be made from assuming a power-law relation $Q(a,\nu)\sim \nu^n$. In such a case,
  \begin{equation}
  T_{\rm dust} = W(r)^{1/(4+n)}\,T_\star,
  \end{equation}
  which for $r\gg R_\star$ becomes
  \begin{equation}
  T_{\rm dust} \approx \left(\frac{R_\star}{r}\right)^{2/(4+n)}\,T_\star.
  \end{equation}
  Considering amorphous carbon grains,
  a reasonable choice for the power index is $n\approx 1$, i.e. a linear dependence on frequency. For the
  strictly linear case we get $T_{\rm dust} \sim T_\star\,r^{-2/5}$, which is consistent with the temperature
  structure found by SS2000 using their DEXCEL-code.
  Hence, we have replaced the energy equation in our CSE model by inserting this approximate temperature structure 
  for the gas temperature
  \begin{equation}
  T_{\rm CSE} \approx T_0\, \left(\frac{r}{R_0} \right)^{-2/5},
  \end{equation}
  where $T_0 \approx 500$ K and $R_0 = 7 \cdot 10^{14} $ cm. This approxmation is what we refer to as Case
  II, which is nearly the opposite to Case I, since the temperature structure is only reflecting the stellar
  radiation field and neither affected by shocks, nor by the expansion. It is reasonable to assume that
  $T_{\rm ad}\le T_{\rm gas}\le T_{\rm dust}$, i.e., the true gas temperature structre will be somewhere in 
  between the temperature structures implied by Case I and Case II.

\section{Results}
  \begin{table*}
  \caption{\label{parameters} Input parameters ($L_\star$, $T_{\rm eff}$, C/O, $P$, $\Delta v_{\rm p}$) and the 
  resulting mean mass loss rate, mean velocity at the outer boundary and mean degree of dust condensation at the outer boundary. 
  $R_\star$ is the stellar radius of the hydrostatic initial model (obtained from $L_\star$ and $T_{\rm eff}$) and $\log(g)$ 
  is the corresponding surface gravity. The dust-to-gas mass ratio $\rho_{\rm dust}/\rho_{\rm gas}$ is calculated from $f_{\rm c}$ 
  as described in H\"ofner \& Dorfi (1997). All models have a mass of $M_\star=1.2M_\odot$ and $\log(\varepsilon_{\rm O}) = -3.35$. 
  The mean values of the wind properties in this table are all taken over at least 100 pulsation periods, which is
  sufficient to ensure a small error due intermittent variations of these quantities.}
  \center
  \begin{tabular}{lcccccccccccc}  
  \hline
  \\
  Model & $L_\star$ & $T_\mathrm{eff}$ & $R_\star$ &
  $\log(g)$ & C/O & $P$ & $\Delta v_{\rm p}$ & $\langle\dot{M}\rangle$ & $\langle v_{\rm ext} \rangle$ & $\langle {\rm f_c} \rangle$ & 
  ${\rho_{\rm dust}\over\rho_{\rm gas}}$\\[2mm]
  & [$L_{\odot}$] &  [$\mbox{K}$] &  [$R_{\odot}$] & & & [days] & [km s$^{-1}$] & [$M_\odot$ yr$^{-1}$] & [km  s$^{-1}$] & & [$10^{-4}$] \\
  \\
  \hline
  \\
  l58t305c201u2 & $5800$  & $3050$ & $343$ & $-0.55$ & $2.01$ & $333$ & $2.0$ & $-$ & $-$ & $-$ & $-$ \\
  l58t305c201u4 & $5800$  & $3050$ & $343$ & $-0.55$ & $2.01$ & $333$ & $4.0$ & $-$ & $-$ & $-$ & $-$ \\
  l58t305c201u6 & $5800$  & $3050$ & $343$ & $-0.55$ & $2.01$ & $333$ & $6.0$ & $-$ & $-$ & $-$ & $-$ \\
  \\
  l80t28c185u2 & $8000$  & $2800$ & $485$ & $-0.85$ & $1.85$ & $436$ & $2.0$ & $-$ & $-$ & $-$ & $-$ \\
  l80t28c185u4 & $8000$  & $2800$ & $485$ & $-0.85$ & $1.85$ & $436$ & $4.0$ & $6.0\cdot 10^{-7}$ & $10$ & $0.16$ & $3.9$\\
  l80t28c185u6 & $8000$  & $2800$ & $485$ & $-0.85$ & $1.85$ & $436$ & $6.0$ & $3.8\cdot 10^{-6}$ & $21$ & $0.38$ & $9.1$\\
  \\
  l80t28c201u2 & $8000$  & $2800$ & $485$ & $-0.85$ & $2.01$ & $436$ & $2.0$ & $-$ & $-$ & $-$ & $-$\\
  l80t28c201u4 & $8000$  & $2800$ & $485$ & $-0.85$ & $2.01$ & $436$ & $4.0$ & $1.3\cdot 10^{-6}$ & $15$ & $0.18$ & $10$\\
  l80t28c201u6 & $8000$  & $2800$ & $485$ & $-0.85$ & $2.01$ & $436$ & $6.0$ & $2.8\cdot 10^{-6}$ & $24$ & $0.42$ & $24$\\
  \\
  l90t275c201u2 & $9000$  & $2750$ & $530$ & $-0.93$ & $2.01$ & $481$ & $2.0$ & $-$ & $-$ & $-$ & $-$ \\
  l90t275c201u4 & $9000$  & $2750$ & $530$ & $-0.93$ & $2.01$ & $481$ & $4.0$ & $2.8\cdot 10^{-6}$ & $21$ & $0.31$ & $18$\\
  l90t275c201u6 & $9000$  & $2750$ & $530$ & $-0.93$ & $2.01$ & $481$ & $6.0$ & $4.8\cdot 10^{-6}$ & $24$ & $0.42$ & $24$\\
  \\
  l105t27c201u2 & $10500$  & $2700$ & $654$ & $-1.11$ & $2.01$ & $547$ & $2.0$ & $1.1\cdot 10^{-6}$ & $16$ & $0.15$ & $8.5$\\
  l105t27c201u4 & $10500$  & $2700$ & $654$ & $-1.11$ & $2.01$ & $547$ & $4.0$ & $4.6\cdot 10^{-6}$ & $17$ & $0.24$ & $14$\\
  l105t27c201u6 & $10500$  & $2700$ & $654$ & $-1.11$ & $2.01$ & $547$ & $6.0$ & $6.7\cdot 10^{-6}$ & $20$ & $0.34$ & $19$\\
  \\
  l12t265c201u2 & $12000$  & $2650$ & $682$ & $-1.15$ & $2.01$ & $611$ & $2.0$ & $3.0\cdot 10^{-6}$ & $19$ & $0.17$ & $9.7$\\
  l12t265c201u4 & $12000$  & $2650$ & $682$ & $-1.15$ & $2.01$ & $611$ & $4.0$ & $7.1\cdot 10^{-6}$ & $18$ & $0.24$ & $14$\\
  l12t265c201u6 & $12000$  & $2650$ & $682$ & $-1.15$ & $2.01$ & $611$ & $6.0$ & $9.9\cdot 10^{-6}$ & $19$ & $0.32$ & $18$\\
  \\
  \hline
  \\
  \end{tabular}
  \label{models}
  \end{table*}

\begin{figure*}[!ttt]
  {\includegraphics[width=18cm]{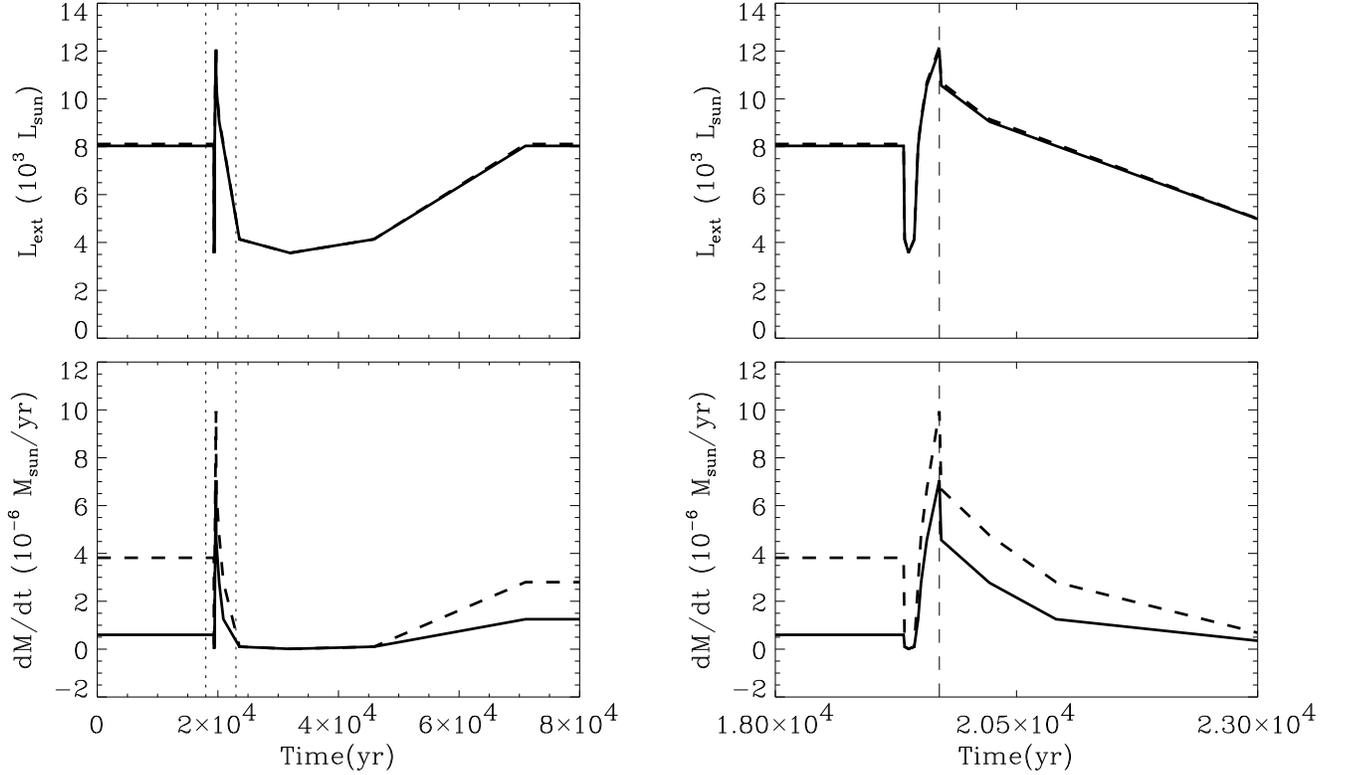}}
  \caption{  \label{timeline}
  Upper left panel: the external luminosity for $\Delta v_{\rm p}=4.0$ kms$^{-1}$ (full line) and $\Delta v_{\rm p}=6.0$ kms$^{-1}$ (dashed line) as a function of time.
  Lower left panel: the mass loss rate for $\Delta v_{\rm p}=4.0$ kms$^{-1}$ (full line) and $\Delta v_{\rm p}=6.0$ kms$^{-1}$ (dashed line) as a function of time.
  The right panels shows a close ups (the interval between the vertical dotted lines) demonstrating the effects of the thermal pulse. The vertical dashed line
  marks the luminosity peak of the He-flash.}
  \end{figure*}

\begin{figure*}[!ttt]
  {\includegraphics[width=18cm]{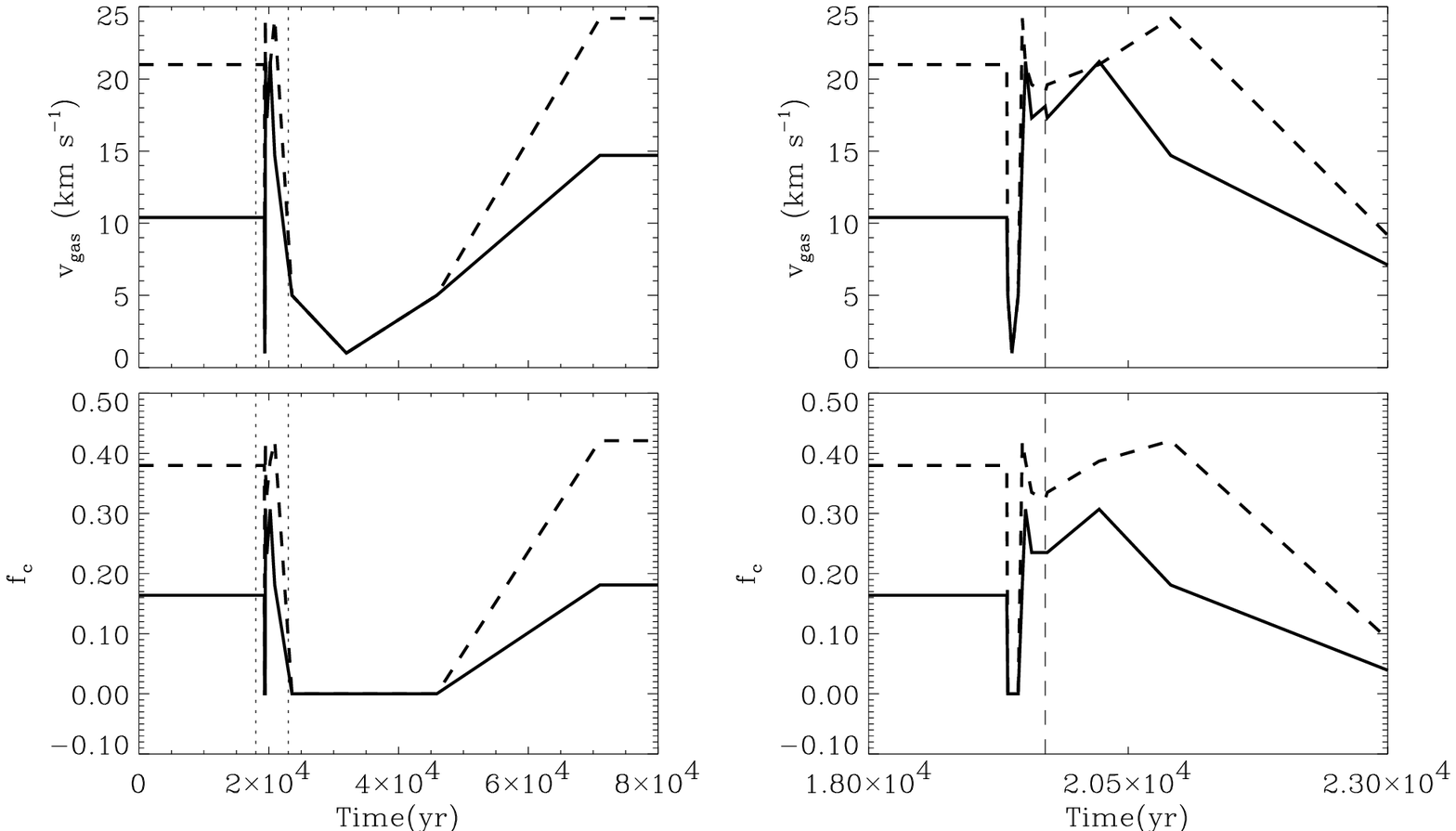}}
  \caption{  \label{timeline2}
  Upper left panel: the wind velocity for $\Delta v_{\rm p}=4.0$ kms$^{-1}$ (full line) and $\Delta v_{\rm p}=6.0$ kms$^{-1}$ (dashed line) as a function of time.
  Lower left panel: the mean degree of dust condensation for $\Delta v_{\rm p}=4.0$ kms$^{-1}$ (full line) and $\Delta v_{\rm p}=6.0$ kms$^{-1}$ (dashed line) as a function of time.
  The right panels shows a close ups (the interval between the vertical dotted lines) demonstrating the effects of the thermal pulse. The vertical dashed line
  marks the luminosity peak of the He-flash.}
  \end{figure*}

  Stellar parameters and model results for the set of atmosphere models used to compute the mass loss history are presented
  in Table \ref{parameters}. A graphical display of the time evolution of the luminosity and the mass loss rate are shown 
  in Fig. \ref{timeline}, where the expected correlation between these two quantities is evident. In Fig. \ref{timeline2} 
  we compare the wind velocity and the mean degree of dust condensation -- two quantities that show a correlation as well. 
  In these plots, full lines represent the case where $\Delta v_{\rm p} = 4.0\,{\rm km\,s^{-1}}$, while dashed lines 
  correspond to $\Delta v_{\rm p} = 6.0\,{\rm km\,s^{-1}}$. Increasing this piston amplitude leads to a much stronger 
  pre-flash wind. The mass loss rate as well as the wind velocity changes significantly (as is evident from Fig. 
  \ref{timeline} and Fig. \ref{timeline2}).
 
  A close up of the evolution during the flash event (right panels in Fig. \ref{timeline} and Fig. \ref{timeline2}) reveals 
  an anti-correlation between mass loss rate and wind velocity at the peak of the flash. This may seem curious, but can be 
  explained by the fact that the relative abundance of dust is lower in the flash-peak model, i.e the model corresponding
  to the highest luminosity (see Table \ref{parameters}). A smaller abundance of dust (lower $\rho_{\rm dust}$) will result
  in a less effective momentum transfer between radiation and matter. Hence, we could expect a lower wind velocity, 
  but not necesserily a lower mass loss rate, since the latter is proportional to both the matter density of the atmosphere
  and the wind velocity. In the particular case of the flash-peak model the gas density is higher, which leads to a higher
  mass loss rate although the wind is slower.
  The correlation between the mean degree of dust condensation $f_{\rm c}$ and the mean wind velocity is seen in Fig. 
  \ref{timeline2}.

  The fact that the mass loss rate, wind velocity and $f_{\rm c}$ are strongly dependent on the piston amplitude 
  $\Delta v_{\rm p}$ means that the predicted wind and mass loss cannot be very precisely constrained. There is no way of 
  knowing if this amplitude can be regarded as constant over time without a complete physical description of the pulsation 
  mechanism. This is certainly an important remark in the present case, since it appears quite likely that the pulsation 
  amplitude could change as the star goes through a He-shell flash. Furthermore, the difference in kinetic energy 
  injection by the pulsations depending on the choice of $\Delta v_{\rm p}$ may have significant effects on the amount of
  material being transferred out to the dust-formation region, and thus also for the degree of dust condensation and 
  consequently for the momentum transfer and wind velocity. For a further discussion on constraints on 
  $\Delta v_{\rm p}$, see Section \ref{mlrdis}.

  \subsection{Time Evolution of the Atmosphere}
  Since the wind properties may show short-term, intermittent variations over time, the model output is time-averaged. 
  The models usually reach a statistical steady state after $\sim 50$ pulsation periods. An interval of $\sim 100$ periods 
  is sufficient to obtain reliable mean values. A typical wind model is evolved over $150-200$ periods, which is thus 
  sufficient for the output to be statistically stable.
  
\begin{figure*}[!ttt]
  {\includegraphics[width=18cm]{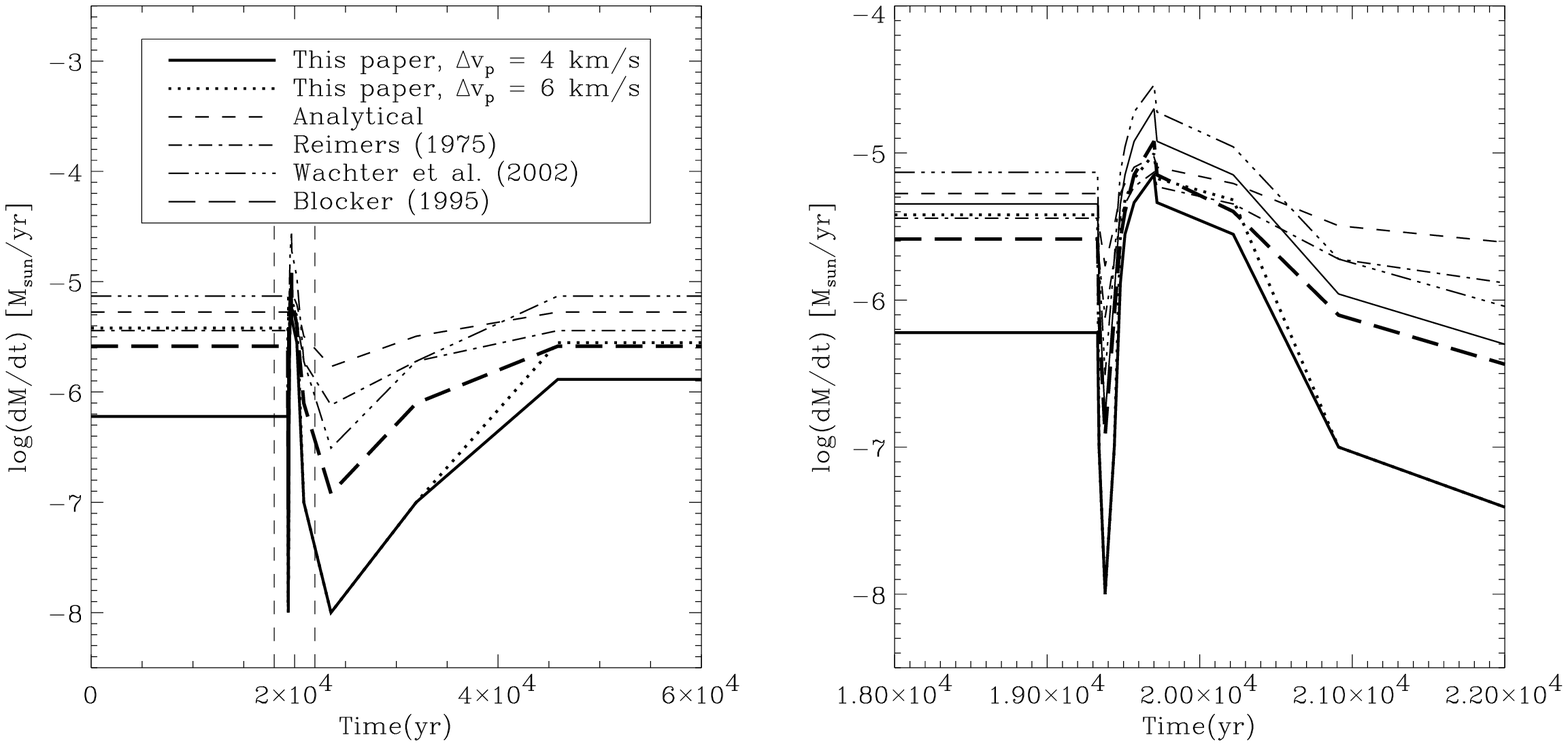}}
  \caption{  \label{mlr_comp}
  Comparison with some existing mass loss formulae. The plot shows the mass loss evolution during the considered
  He-shell flash adopting mass loss rates from our atmosphere model (for two different values of 
  $\Delta v_{\rm p}$), along with a simple analytic model (\cite{Lamers99}), the empirical Reimers (1975) 
  formula and the mass loss prescriptions derived from grids of numerical models by Wachter et al.  
  (2002) and Bl\"ocker (1995). Due to the large differences we have plotted the \it logarithm \rm of the mass loss rate.
  The right panel shows a blow-up of the interval marked with vertical dashed lines in the left panel. Note as well,
  that for the evolution computed in this paper, the values at the mass loss minimum are {\it ad hoc} values.}
  \end{figure*}

  There is a complex interplay between density, dust formation, radiation pressure and gravity (escape velocity) 
  that makes the variations in mass loss for different sets of stellar parameters highly nonlinear. This interplay
  is why simple parameterisations, where the mass loss rate is essentially just a function of the luminosity (e.g. 
  \cite{Blocker95}, \cite{Wachter02}), cannot provide a satisfactory prescription for mass loss throughout the evolution of
  a star on the AGB. The well-known (but unfortunately often abused) Reimers formula (\cite{Reimers75}) provides a mass 
  loss prediction that cannot be made in agreement with the resulting mass loss evolution from our models. Regardless 
  of the choice of the parameter $\eta_{\rm R}$ (in Fig. \ref{mlr_comp} the case $\eta_{\rm R}= 1$ is shown) the mass loss
  evolution is not as strongly peaked at the time of the He-flash as the evolution implied by our models. However, Reimers 
  formula was intended as a description of the mass loss from red giants, which does not include TP-AGB stars, and there are 
  no justifications for extending the use of Reimers formula to this regime. Its linear dependence on the luminosity 
  $L_\star$ means that it cannot at the same time reproduce the mass loss associated with the flash peak. A comparison of 
  our results with existing mass loss formulae is presented Fig. \ref{mlr_comp}, in which it becomes clear 
  that none of the simple descriptions considered here agrees with our detailed model, nor are they consistent with each 
  other. It is particularly interesting to notice that the description by Wachter et al. (2002) yields a significantly 
  higher over-all mass loss rate compared with other prescriptions, as well as our wind model. The reason for this, as far 
  as we can tell, is two-fold: Wachter et al. use grey radiative transfer (with lower gas opacities) which is known to give 
  higher mass loss rates and they derive their formula in principle only from models with saturated dust-formation, i.e. models 
  in the regime where the momentum transfer is most efficient.

  In both Case I and Case II, this thin shell structure survives over the whole simulated CSE evolution. In Case 
  II the thin shell structure remains quite intact as the shell expands -- the only important change over time is a slight 
  broadening of the density peak likely due to internal pressure. The evolution of Case I is somewhat different. Here a 
  very pointed shell structure evolves shortly after flash event, but as the shell expands two shock fronts are formed, 
  creating a density bump (or plateau) altough the initial density peak remains. The density peak is located at 
  the contact discontinuity between the slow (pre-flash) and fast wind phases, where the gas temperature reaches a minimum.
 
  \subsection{Formation of a Detached Shell}
  \label{cseres}
  The results of the hydrodynamic CSE evolution modelling confirm the "modified He-shell flash scenario" presented by 
  SS2000 in the case of $\Delta v_{\rm p}=4.0$ km s$^{-1}$. Using the mass loss evolution we have obtained from 
  our detailed modelling, we receive a density structure that is qualitatively consistent with the results of SS2000, 
  but in general results in a much thinner shell. Case I also shows a time variation of the relative
  thickness not present in the results of SS2000 (see Fig. \ref{thickness}, Fig. \ref{fixedtemp} in this paper and Fig. 12 
  in SS2000). The reason for this time variation of the thickness is probably the different evolution of the velocity 
  profile shown in our results, shortly after the He-shell flash. As the CSE evolves, the velocity profile becomes very 
  similar to the result from the "modified He-shell flash scenario" presented in SS2000.

  We have also calculated the FWHM of the quantity, $dm/dr= 4\pi\,r^2\rho$, which we will refer to as the 
  thickness of the shell. In Fig. (\ref{thickness}) we show the time evolution of the thickness of the shell for Case I and 
  II. As is evident from the plots, including a reasonable temperature structure has really a dramatic effect on the CSE 
  evolution. Beyond $R = 10^{17}$ cm (roughly $2000$ years after the flash event) the relative thickness 
  $\Delta R_{\rm s}/R_{\rm s}$ of the shell appears constant in Case II, while in Case I we see more of a 
  $\Delta R_{\rm s}/R_{\rm s} \sim 1/R_{\rm s}$ behaviour. The typical relative thickness in Case II is 
  $\Delta R_{\rm s}/R_{\rm s} \sim 0.01$, which is less than what has been derived from observations (e.g. 
  \cite{Olofsson00}). It should be pointed out, however, that these observations of detached shells we refer to are CO 
  detections. Gonz\'alez-Delgado et al. (2001) \nocite{Gonzalez-Delgado01} have detected shell structures around AGB stars 
  in the optical, which have $\Delta R_{\rm s}/R_{\rm s} \sim 0.1$. This is an order of magnitude thicker than our model 
  result and may indicate a problem with the assumptions made in Case II. In Case I the thickness is measured on the very 
  sharp and pointed density peak (making the shell appear extremely thin), which means that the numbers we derive are only 
  of interest for comparison with Case II. 

  The differences ocurring from the use of boundary conditions derived from atmosphere models with different piston 
  amplitudes, are quite dramatic. With $\Delta v_{\rm p}=4.0$ km s$^{-1}$ the CSE model provides a pronounced shell structure,
  while the $\Delta v_{\rm p}=6.0$ km s$^{-1}$ case does not lead to the formation of any shell structure at all (see Fig. 
  \ref{norad} and Fig. \ref{fixedtemp}). Since the $\Delta v_{\rm p}=2.0$ km s$^{-1}$ case was directly excluded (no wind, 
  except at the flash peak) we are left with only one case that can produce a detached shell. In this case the density 
  contrast at the shell radius is roughly two orders of magnitude relative to the surrounding medium, which should 
  definitely be a detectable feature. 

  In order to determine if there are any effects on the shell formation from the short term variations of the wind during 
  the flash (see Fig. \ref{timeline2}), we have integrated the mass loss during the flash event and created a representative 
  pulse function (same integrated mass loss) and replaced the time dependent inner boundary condition with this simplified 
  function (see Fig. \ref{bc_box}). The wind velocity is turned into a pulse function in the same fashion. Comparing the 
  resulting CSE evolution with 
  "case B" and "case C" in SS2000, it is clear that our CSE model produces a thinner and less massive shell, but still with an 
  equally large density contrast. It is also evident that the small features (e.g. the dip at the flash peak) in the wind 
  evolution found in our detailed modelling, have very little effect on the properties and evolution of the CSE and the 
  detached shell (compare Fig. \ref{norad} and Fig. \ref{fixedtemp} with Fig. \ref{norad2} and Fig. \ref{fixedtemp2}). 

  For the sake of completeness, we have tested how the mass loss history suggested by observations of TT Cygni 
  (\cite{Olofsson98}, see also Fig. 1 and Fig. 4 in SS2000) affects the formation of a detached shell, using a boundary for
  the wind velocity which is equivalent to "case B" and "case C" in SS2000. As we show in Appendix \ref{app}, the results 
  we get for Case II are fully consistent with the results of SS2000 using their "NEBEL" code. 

  These tests elucidate the importance of modelling the wind evolution correctly. If the detailed inner boundary conditions
  derived from wind models are replaced by a combination of simple pulse functions (conserving differences in mass loss rate 
  and wind velocity, before, during and after the flash), the CSE evolution is not much affected. If, on the other hand, 
  the mass loss and velocity evolution differs as in "case B" and "case C" in SS2000, the CSE evolution will be different 
  and results in a much thicker shell structure (e.g., compare Fig. \ref{fixedtemp} with Fig. \ref{testcaseB2} and Fig. 
  \ref{testcaseC2}). Thus, it is very important to use consistent combinations of mass loss rates and wind velocities when 
  modelling the evolution of the CSE.

  \begin{figure}
  \resizebox{\hsize}{!}{\includegraphics[width=5.7cm]{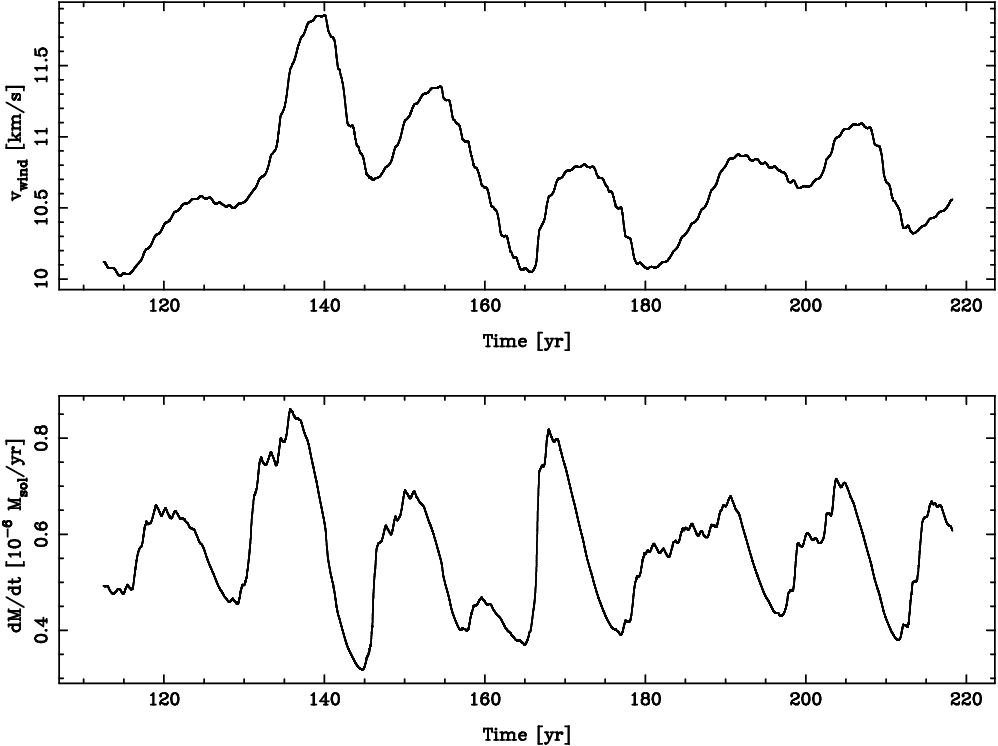}}
  \caption{  \label{timeseries1}
  Timeseries of the wind velocity and mass loss rate at $R_0$ showing the short-term variations before the 
  He-flash (model l80t28m12c185).}
  \end{figure}

  \begin{figure}
  \resizebox{\hsize}{!}{\includegraphics[width=5.7cm]{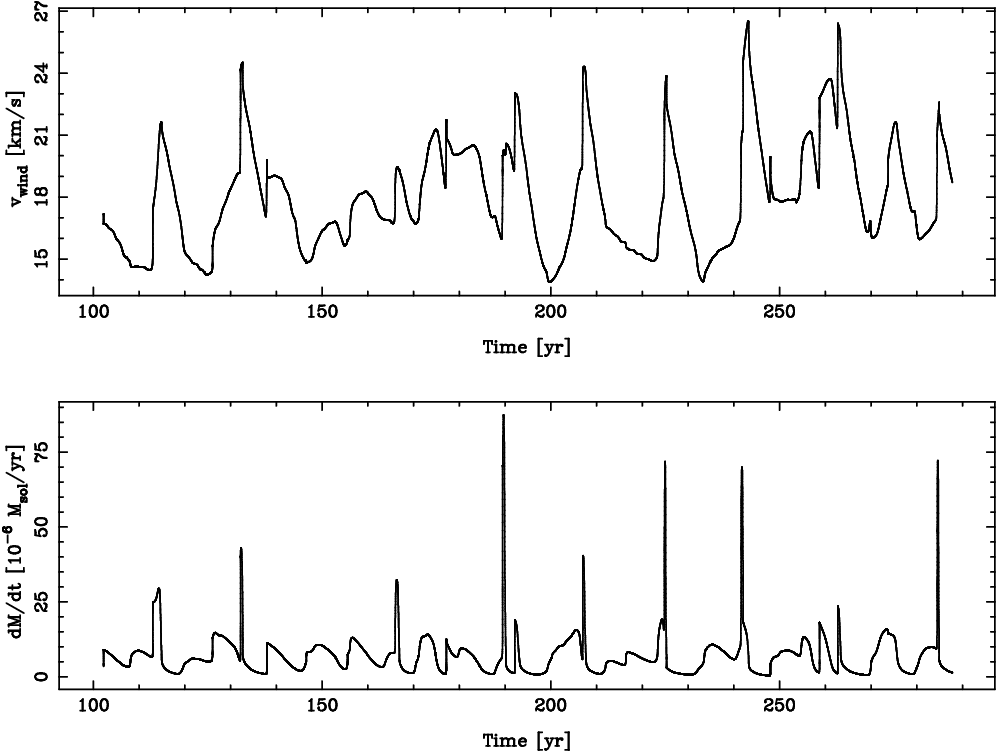}}
  \caption{  \label{timeseries2}
  Timeseries of the wind velocity and mass loss rate at $R_0$ showing short-term variations {\it during the 
  peak of the He-flash} (model l12t265m12c201).}
  \end{figure}

  \begin{figure}
  \resizebox{\hsize}{!}{\includegraphics[width=5.7cm]{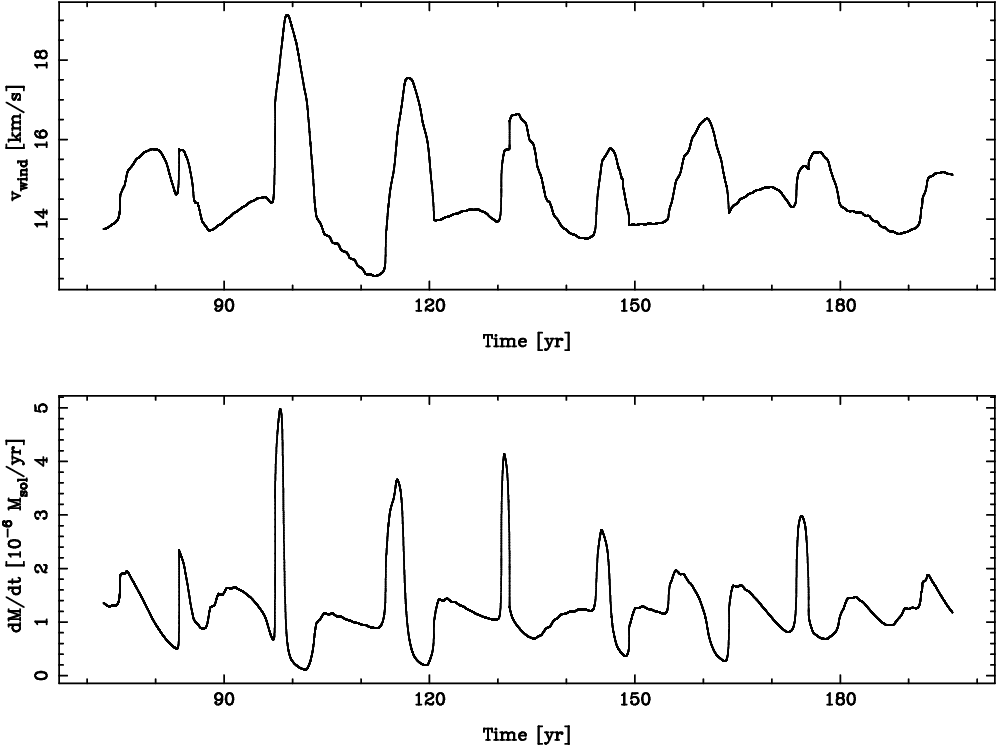}}
  \caption{  \label{timeseries3}
  Timeseries of the wind velocity and mass loss rate at $R_0$ showing short-term variations after the 
  He-flash (model l80t28m12c201).}
  \end{figure}

\section{Discussion}

\subsection{Mass Loss Rates}
  \label{mlrdis}
  The pulsational instability in the stellar interior provides a mechanism for levitation of material into the outer 
  atmosphere, i.e. the injection of kinetic energy from the pulsation related shock fronts as they propagate outwards may 
  be sufficient to push material out to a radius where dust can begin to form. It is this phenomenon 
  that enables the large mass loss rates associated with TP-AGB stars. 

  As described in Section \ref{dynamic} the pulsation is simulated by applying a piston boundary condition. Increasing 
  the piston amplitude in the models from 
  $\Delta v_{\rm p} = 2.0\,{\rm km\,s^{-1}}$ to $\Delta v_{\rm p} = 6.0\,{\rm km\,s^{-1}}$ can raise the density in the 
  dust forming region by as much as one order of magnitude. Consequently, the amount of dust that forms may increase 
  significantly as well and in turn lead to a more efficient momentum transfer and thus greater wind velocities. One can
  in princple fine tune $\Delta v_{\rm p}$ in order to get the wind properties that lead to shell formation for a specific 
  set of stellar parameters. However, even if it is difficult to put strong 
  constraints on the mass loss evolution without a reliable model for the pulsation mechanism, we can still limit the 
  range of piston amplitudes that are reasonable in the present context. As we have seen, too small amplitudes do not lead 
  to wind formation (except at the flash peak) and too large amplitudes will not provide the pronounced variations of wind 
  properties needed to obtain the interacting wind phases that are crucial for shell formation. Hence, the possibility to 
  choose the piston amplitude is restricted and we have found that the amplitude implied by previous comparsions with 
  observations ($\Delta v_{\rm p} = 4.0$ km s$^{-1}$) is a good choice (e.g. \cite{Nowotny05b}). Furthermore, models with 
  this amplitude give mass loss rates and wind velocities that matches observations of TP-AGB stars with known detached 
  shells very well (e.g. \cite{Schoier05}). From recent observations of all objects with known detached molecular shells, 
  Sch\"oier et al. (2005) derived a post-flash mass loss rate (which at the same time gives a hint about the pre-flash rate)
  of a few times $10^{-7}M_\odot$ yr$^{-1}$, and $\sim 10^{-5}M_\odot$ yr$^{-1}$ associated with a flash event, which is 
  almost exactly what we get (see Fig. \ref{timeline}). The picture one gets is rather consistent -- the pulsations are 
  nicely modelled by a piston boundary condition with $\Delta v_{\rm p} \sim 4.0$ km s$^{-1}$ even if the physical 
  mechanism is not modelled. We have thus not considered temporal variations of the pulsational amplitude during the flash 
  event, even if they may occur.

  \begin{figure}
  \includegraphics[width=8.5cm]{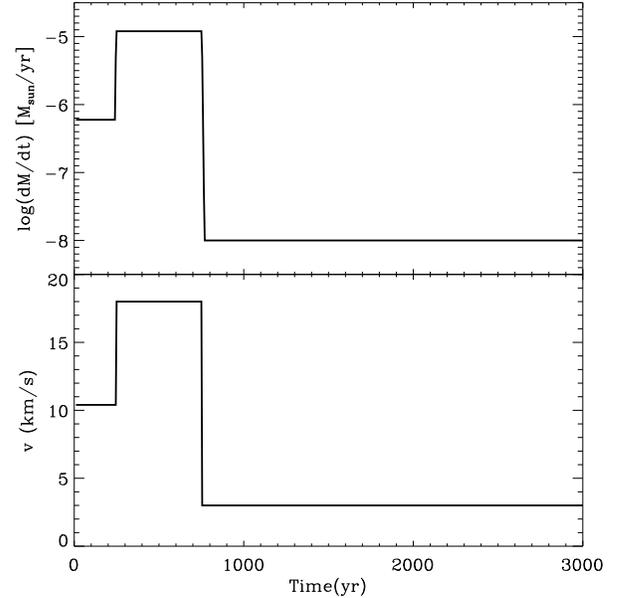}
  \caption{Plots of the pulse functions used to replace the time dependent inner boundary condition. \label{bc_box}}
  \end{figure}

  The three types of models that we have combined provide a rather consistent connection between a He-shell flash and
  detached shells with essentially one adjustable parameter, $\Delta v_{\rm p}$. However, there is one possible inconsistency 
  that we have not mentioned so far. The evolutionary tracks for the present stellar model have been calculated using the 
  mass loss prescription provided by Bl\"ocker (1995). As  shown in Fig. \ref{mlr_comp}, the mass loss rates predicted by 
  our model differ from most simple mass loss prescriptions (formulae). A significantly different mass loss evolution will 
  alter the late stages of stellar evolution (as pointed out by \cite{Blocker95}) and thus provide a different set of stellar 
  parameters as input to our model of the wind evolution. But this is most likely a minor problemns in terms of consistency,
  since the time interval we model is short on an evolutionary time scale and each one in the series of thermal pulses 
  corresponds to slightly different sets of parameters, regardless of how the mass loss affects the stellar evolution. 
  Moreover, the flash event itself, is not regulated by mass loss but by nuclear reaction rates. In our modelling we 
  have picked one typical flash event that looked suitable in terms of getting a reasonable pre-flash mass loss rate, 
  judging from previous experience of mass loss modelling.

\subsection{CSE Evolution and the Formation of a Detached Shell}
  Let us now discuss how a detached shell may, or may not, be formed in the two limit cases of our CSE model (see Section 
  \ref{cse_evol}) due to the wind evolution obtained from our wind models.

  In Case I, where the global temperature structure is set by the expansion of the gas, the shock is basically adiabatic. 
  The increase of the pressure leads to high temperatures (typically a few 
  $10^3$ K) in the shocked regions, which cannot be radiated away (since radiation is not included) and thus enforces a 
  change in the gas motion that manifests itself by the build up of a shock 
  wave in front of the ejected shell. This is why the shell evolves into a density bump with a density peak on top of it. 
  There are several arguments against this kind of "adiabatic" evolution of the shell, where the most important one is that
  any shock front of this type must lose at least some energy through radiation. The high temperature we obtain 
  inside the shocked regions as well as the low temperatures in the stationary pre-flash wind, indicate that Case I is 
  quite unrealistic. 

  \begin{figure*}[!ttt]
  {\includegraphics[width=18cm]{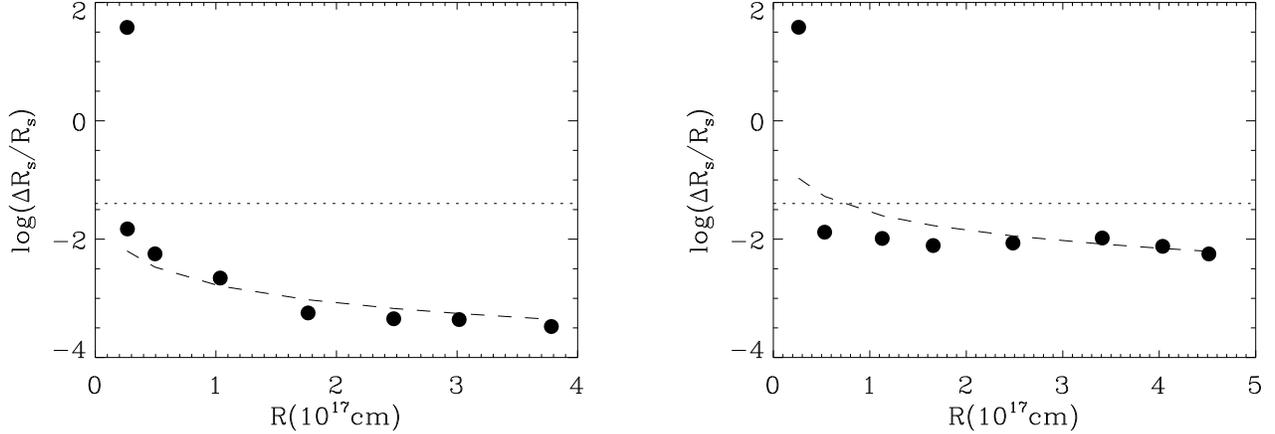}}
  \caption{\label{thickness} Thickness of the detached shells formed in Case I \& II. 
  The left panel shows Case I and the right panel shows Case II for $\Delta v_{\rm p}=4.0$ kms$^{-1}$. 
  The dashed lines indicate fits to $\Delta R_{\rm s}/R_{\rm s} \propto 1/R_{\rm s}$. Each black dot represents
  a specific time in the evolution of the detached shell in accordance with all following plots of the velocity and density
  evolution. The dotted line marks the lower limit for the typical shell thickness derived by Olofsson et al. (2000) from CO
  observations.}
  \end{figure*}

  Case II, where we have assumed a fixed temperature structure, corresponds to having an isothermal shock front relative to
  its surroundings. This is in principle the opposite extreme of Case I, since the expanding shell is forced to have the 
  same temperature as the surrounding medium, meaning that all excess energy is transferred away from the shock. Such an 
  effectiveness in radiative cooling is quite unlikely, but Case II is probably still closer to reality than Case I -- 
  especially regarding the global temperature structure. To make a truly realistic model including all possible energy 
  sinks and sources in this case, it is probably necessary to include a full RHD description (with dust) of the problem 
  and maybe even go from 1D to 3D in order to correctly treat turbulence and dynamical instabilities (see \cite{Myasnikov00}).   

  The reverse shock and the forward shock in front of the shell are effects due to the 
  quite large velocity jumps that build up on each side of the shell as it propagates outward. This feature is most prominent 
  in Case I, due to the overestimation of compressional heating and the resulting thermal pressure gradients.

  As we pointed out in Section \ref{cseres}, changing the mass loss and wind velocity evolution into simple step functions 
  (where the integrals are preserved) does not notably affect the resulting shell structure. The reason is that the 
  formation of a thin shell happens very much through wind interaction, and since we do not change the wind
  velocity of the pre-flash phase or at the mass loss maximum, the wind interaction takes place in a similar fashion. 

  We return now to the importance of having a correct wind model as the basis for the boundary variations in the CSE
  model. In SS2000 the inner boundary of the atmosphere model is prescribed such that a mass loss rate is assumed and the 
  computations start beyond the condensation radius where the wind velocity is set to some initial value (see 
  \cite{Steffen98} for details). The velocity field is thus not derived from a fully self-consistent description of the 
  stellar wind formation (although the equation of motion is solved, subject to the given boundary conditions). Moreover, 
  the mass loss rate is not computed, but rather chosen according to a semi-empirical mass loss formula (\cite{Blocker95}).
  For the CSE evolution to be correcly modelled, 
  however, we must be certain that the variations of both the mass loss rate and the wind velocity are essentially 
  correct and consistent, since the interaction of wind phases is perhaps the most important ingredient. 

  \begin{figure*}[!ttt]
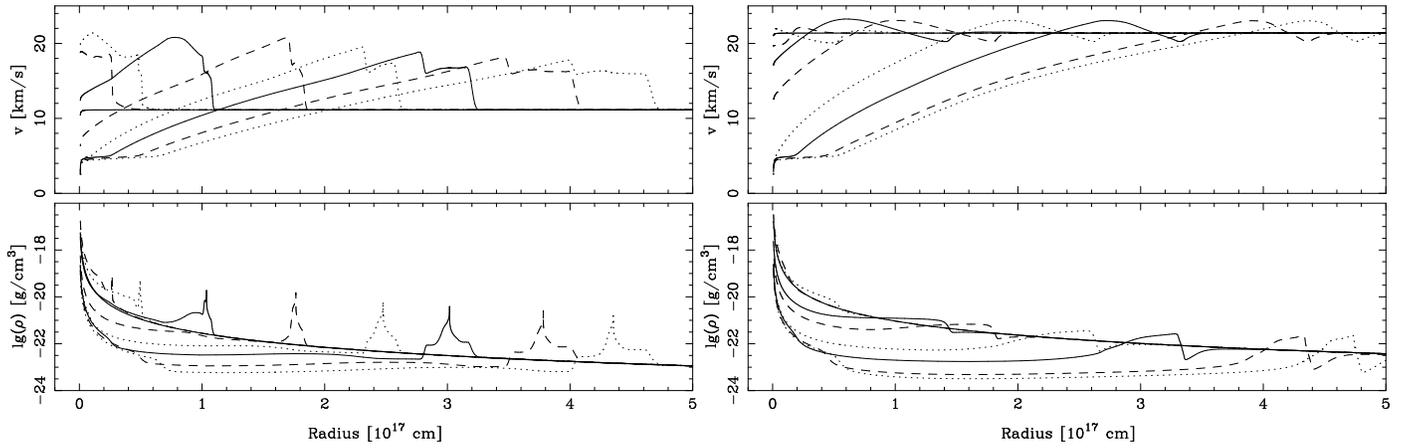

  \resizebox{\hsize}{!}{\includegraphics[angle=270]{6368f10a}
  \includegraphics[angle=270]{6368f10b}}
  \caption{  \label{norad}
  Nine instants in the evolution of the CSE.
  Velocity and density structure of the CSE in Case I.
  Upper panels: the velocity field for $\Delta v_{\rm p}=4.0$ kms$^{-1}$ (left) and $\Delta v_{\rm p}=6.0$ kms$^{-1}$ 
  (right).
  Lower panels: the logarithmic density for $\Delta v_{\rm p}=4.0$ kms$^{-1}$ (left) and 
  $\Delta v_{\rm p}=6.0$ kms$^{-1}$ (right).}
  \end{figure*}

  \begin{figure*}[!ttt]
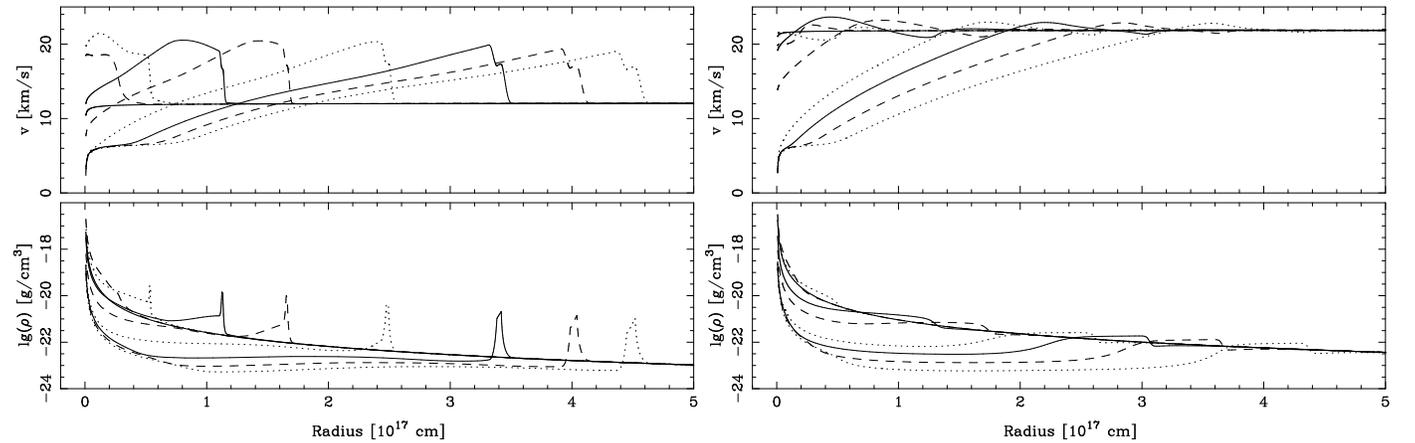

  \resizebox{\hsize}{!}{
  \includegraphics[width=5.7cm, angle=270]{6368f11a}
  \includegraphics[width=5.7cm, angle=270]{6368f11b}}
  \caption{  \label{fixedtemp}
  Same as Fig. \ref{norad}, but for Case II.}
  \end{figure*}

  Our detailed wind models (for $\Delta v_{\rm p}=4.0$ km s$^{-1}$) show that the wind velocity evolves from a slow phase 
  ($\sim 10$ km s$^{-1}$) before the He-shell flash, into a rather fast wind ($\sim 20$ km s$^{-1}$) during He flash event 
  and then back to a slower wind phase again (see Fig. \ref{timeline}). This is largely the same type of evolution used in
  the "modified He-shell flash scenario" in SS2000, although we have obtained it using a more sophisticated model, which 
  provides both mass lass rate and wind velocity as functions of stellar parameters. 

  In the $\Delta v_{\rm p} = 6.0$ km s$^{-1}$ case we obtain a very different evolution compared with the 
  $\Delta v_{\rm p} = 4.0$ km s$^{-1}$ case, also qualitatively speaking. The pre-flash mass loss rate is almost an order 
  of magnitude higher and the wind velocity is a factor of two higher, providing almost no variations in the wind 
  velocity before and during the flash. The latter is the main reason why the $\Delta v_{\rm p} = 6.0$ km s$^{-1}$ case 
  does not render a shell structure. 

  How reliable are these results then? One may
  ask whether, in some part, the pre-flash wind could be different if drift/coupling between the gas and dust phases was 
  to be introduced. The question is very relevant since these effects are treated in SS2000. 
  At the flash peak this is not a problem, since high mass loss rate results in a higher CSE density which makes the 
  effects of drift much less significant. We are likely to overestimate the pre-flash wind velocity due to 
  the phase coupling. Hence, the pre-flash wind velocity would probably be even lower if drift was to be included, making 
  the velocity jump (see Fig. \ref{timeline2}) even larger. Therefore, the inclusion of drift would probably not change the
  result qualitatively.
  
  The thickness $\Delta R_{\rm s}$ of the shell is hard to evaluate in Case I, since the density peak has almost no spatial 
  extension compared with the scales considered here, and the bump is probably a somewhat artificial structure created by 
  a dramatic pressure increase at the shock fronts since no energy is radiated away. It is therefore difficult to interpret
  the meaning of the thickness as defined by the FWHM. In Case II the characteristic absolute thickness of the shell 
  increases almost linearly with the shell radius (which in turn grows close to linearly with time). This results in a 
  constant $\Delta R_{\rm s}/R_{\rm s}$-ratio, in agreement with the result found by SS2000 in the "modified He-shell flash 
  scenario". However, where SS2000 find a ratio $\Delta R_{\rm s}/R_{\rm s} \approx 0.04$, we find this number to be 
  roughly $0.01$, as we have pointed out previously in Section \ref{cseres}. If the faster wind associated with the "mass 
  loss eruption" runs into a less dense medium, as in our model, it means that less matter is swept up and a thinner shell 
  may thus be formed out of this interaction. But as far as we can tell, the thinner shells can not be explained only by 
  the weaker pre-flash wind that we obtain, nor can it be explain by any significant difference in the duration of the
  mass loss eruption. A possible, additional explanation might be the fact that the shocks are 
  isothermal with their surroundings in our model, and therefore the shell structure is not substantially broadened by 
  thermal pressure. But the conditions in the SS2000 model are also close to isothermal, which makes this explanation
  seem less likely. A perhaps more likely explanation arises from the fact that the wind {\it velocity} jump from pre-flash to 
  the flash-peak is an important parameter governing the formation of a detached shell. In the present work this jump is 
  roughly a factor of 2 larger than in SS2000, which may be a viable explanation. In general, however, our results confirm 
  the connection between stellar evolution (the He-shell flash) and the formation of detached geometrically thin shells 
  around TP-AGB stars.

\section{Summary and Conclusions}
  We have used the RHD code for dynamic atmospheres by H\"ofner et al. (2003) in combination with a stellar evolutionary 
  track of a $M = 2 M_\odot$, $Z = 0.01$ star in order to calculate the evolution of mass loss and wind velocity during a
  He-shell flash. The observed wind and mass loss properties associated with detached shells are rather well reproduced by 
  our models of TP-AGB star atmospheres using stellar parameters sampled from this evolutionary track. None of the other 
  existing mass loss prescriptions for AGB stars we have considered give, quantitatively speaking, mass loss rates
  similar to those which we 
  compute. However, constraints on the mass loss evolution derived from observations (\cite{Olofsson00}, \cite{Schoier05}) 
  appear to agree better with our results than with these mass loss formulae. Our numerical model may therefore be 
  considered as a reasonably realistic description of the connection between stellar parameters and mass loss rates for 
  TP-AGB stars. 

  We find that both mass loss rate and wind velocity increase during the flash event. This is in qualitative agreement 
  with the "modified He-shell flash scenario" proposed in SS2000 and we thus conclude that geometrically thin detached
  shells are neither likely to be formed by simple two-wind interaction nor out of a "mass loss eruption" (with no 
  variations in the wind velocity) alone. In fact, the formation of a slow and a fast wind combined with an eruptive mass 
  loss associated with the fast wind turns out to be very critical. \it To use consistent combinations of mass loss rates 
  and wind velocities is therefore most important in order to actually investigate a connection between He-shell flashes 
  and detached shells. \rm The amplitude of the internal pulsations of the star has a great effect on the wind velocities 
  before and after the flash and thus on the strength of the wind interaction, in the case studied here.

  \begin{figure}[!ttt]
  \resizebox{\hsize}{!}{
  \includegraphics[width=5.7cm, angle=270]{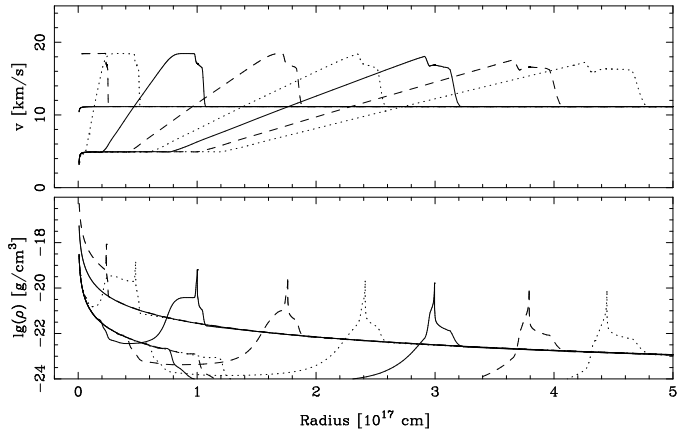}}
  \caption{  \label{norad2}
  Nine instants in the evolution of the CSE.
  Velocity and density structure of the CSE using the step function boundary condition (see Section \ref{cseres})
  in Case I ($\Delta v_{\rm p}=4.0$ kms$^{-1}$). }
  \end{figure}
    
  \begin{figure}[!ttt]
  \resizebox{\hsize}{!}{
  \includegraphics[width=5.7cm, angle=270]{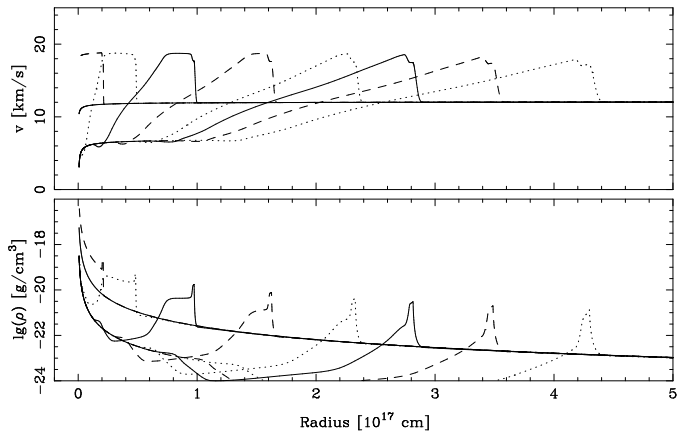}}
  \caption{  \label{fixedtemp2}
  Same as Fig. \ref{norad2}, but for Case II.}
  \end{figure}
  
  Our simplified dynamical model of CSE evolution agrees qualitatively with what is expected from observations. We 
  considered two basically complementary approximations for radiation effects. First we considered the case with no 
  radiation effects at all (Case I) and then a second case where we used a simple analytic model for the temperature 
  structure based on radiative equilibrium for the dust (Case II). In both cases we have assumed that all forces, except thermal 
  pressure, are negligible. Case I and Case II agree qualitatively with each other, in the sense that they both produce a 
  geometrically thin detached shell. However, only in Case II the relative thickness of the shell seems marginally consistent with CO 
  observations. The thin shell structure seems to be well maintained throughout the evolutionary time span of the model, 
  but the thickness of the shell grows linearely as it expands, which is most likely an effect of the internal pressure. 
  Case I initially forms a shell structure with a much more pointed appearence, which in fact is the contact discontinuity 
  at the interface between the winds. The pointed structure seems stable, but as the expansion continues a reverse shock
  propagating inwards is created, as well as an outgoing shock front which sweeps up a large portion of the matter ejected 
  by the pre-flash wind. As a consequence of this, the shell structure evolves into a wide "density bump" with an 
  unreasonably thin shell structure on top of it. From this we conclude that the temperature structure is one of the key 
  components for the shell formation. In addition, simple 1D models may be inadequate since a full 3D simulation may result
  in much more complex structures that, in turn, may have significant effects on the formation and evolution of the shell.
  
  Further work is obviously needed, but so far we can conclude that the He-shell flash and the associated "mass loss 
  eruption" in combination with wind interaction gives a satisfying and consistent description of the formation of detached
  shells around TP-AGB stars. Our detailed modelling of the mass loss evolution shows that a very short, intense period of 
  mass loss can be associated with a He-shell flash. The "mass loss eruption" is followed by a period of no or very little 
  mass loss, before a steady wind phase is once again established. This "eruptive" mass loss evolution does not seem to 
  lend much support for the simple two-wind interaction scenario where a constant faster wind, with a higher mass loss rate 
  associated with it, runs into a slower wind. Since a combination of a "mass loss eruption" and interacting wind phases is
  consistent with both stellar evolution and the observed properties of detached shells, we find this scenario to be the
  most likely. 

  This paper has presented a first attempt to construct a self-consistent model of the formation of detached shells. It is,
  however, important to consider the CSE evolution in more detail than we have done here. A full RHD treatment in 3D would
  probably show additional features (e.g. instabilities) that might help to constrain the mass loss history in more detail.
  Our intention is to return to this issue in a future publication.

\begin{acknowledgements} 
The authors wish to thank the referee, M. Steffen, for his careful reading of the manuscript and constructive criticism that
helped to improve the clarity of this paper.
B. Gustafsson is thanked for his reading of the manuscript draft and many valuable comments on the project. K. Eriksson
and R. Wahlin are thanked for stimulating discussions on AGB stars in general. This work was partly supported by the
Swedish Research Council (Vetenskapsr\aa det). 
F.H. acknowledges support by the LDRD program (20060357ER) at Los Alamos National Laboratory. Part of this work has been carried out in collaboration with the NSF Physics Frontier Center JINA (Joint Insititute for Nuclear Astrophysics).
\end{acknowledgements}

\Online

\begin{appendix}
  \section{Comparison with case "B" and "C" in SS2000}
 \label{app}
  This appendix presents test cases using basically the same boundary variations as in SS2000. Their "NEBEL" code contains 
  a detailed treatment of radiation and cooling, compared with our fixed temperature structure in Case II. The "DEXCEL" 
  runs have no directly corresponding case among the ones we have tried here. But since the "DEXCEL" code is known to 
  suffer from numerical diffusion (which was pointed out in SS2000) any direct comparison would be difficult. Nonetheless, we
  think it is important to show that our code (Case II) produces results that are comparable to the "NEBEL" runs presented
  in SS2000 if the input is similar.  

\begin{figure}
  \resizebox{\hsize}{!}{\includegraphics[width=5.7cm, angle=270]{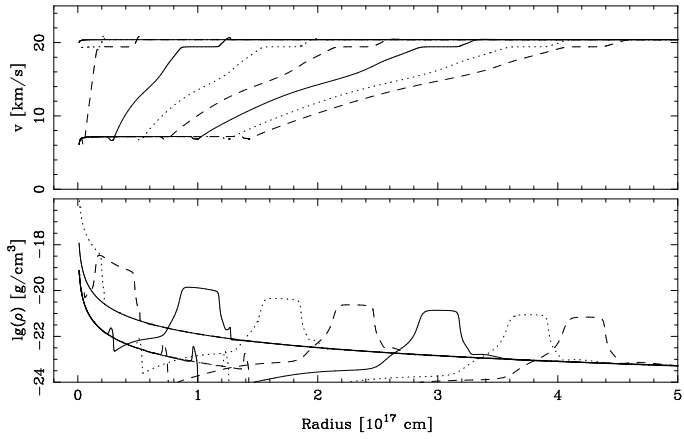}}
  \caption{  \label{testcaseB}
  Nine instants in the evolution of the CSE.
  Upper panel: velocity as function of radius for Case I with the same boundary condition as "case B" in SS2000.
  Lower panel: density as function of radius for Case I with the same boundary condition as "case B" in SS2000.
   }
  \end{figure}

\begin{figure}
  \resizebox{\hsize}{!}{\includegraphics[width=5.7cm, angle=270]{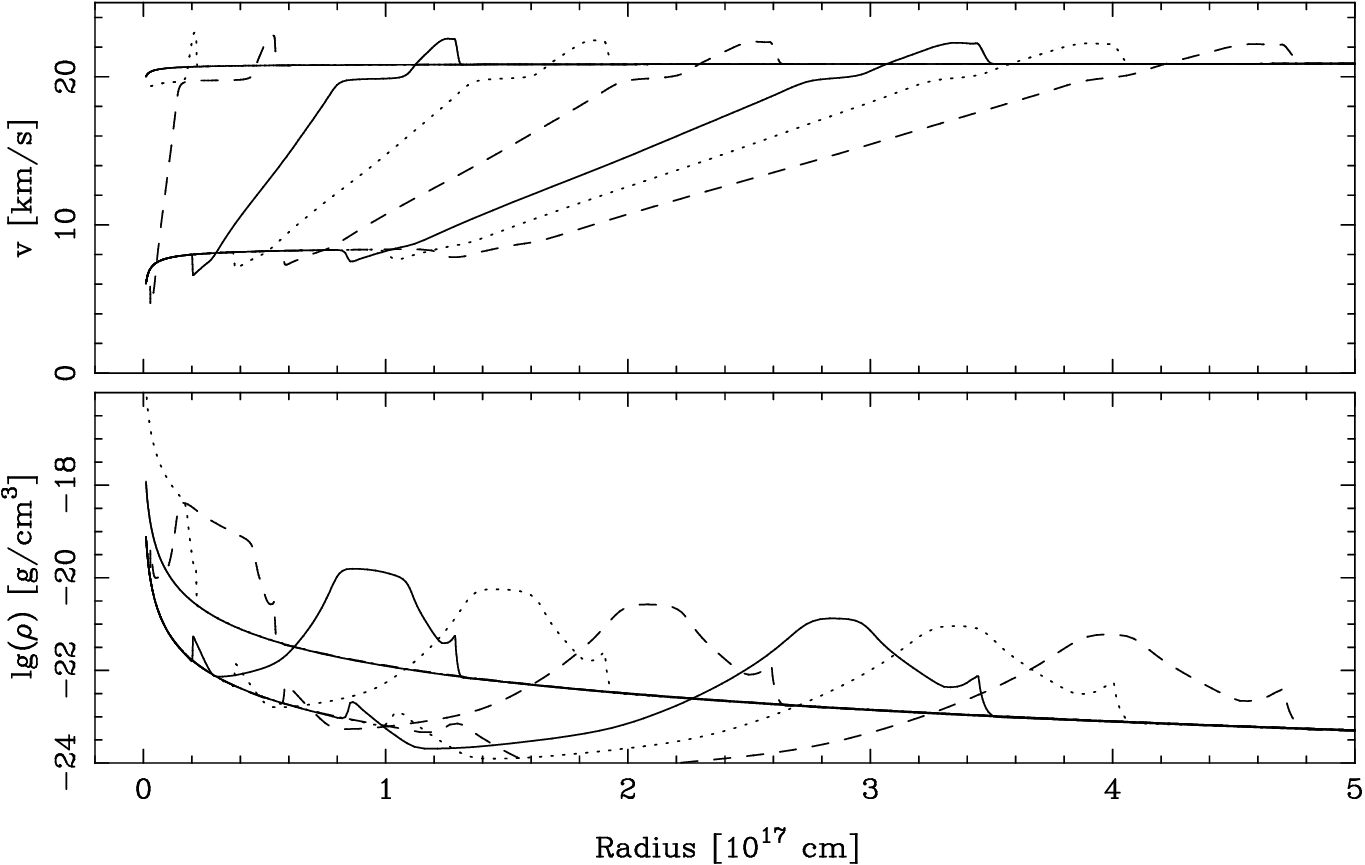}}
  \caption{  \label{testcaseB2}
  Same as Fig. \ref{testcaseB}, but for Case II.}
  \end{figure}

\begin{figure}
  \resizebox{\hsize}{!}{\includegraphics[width=5.7cm, angle=270]{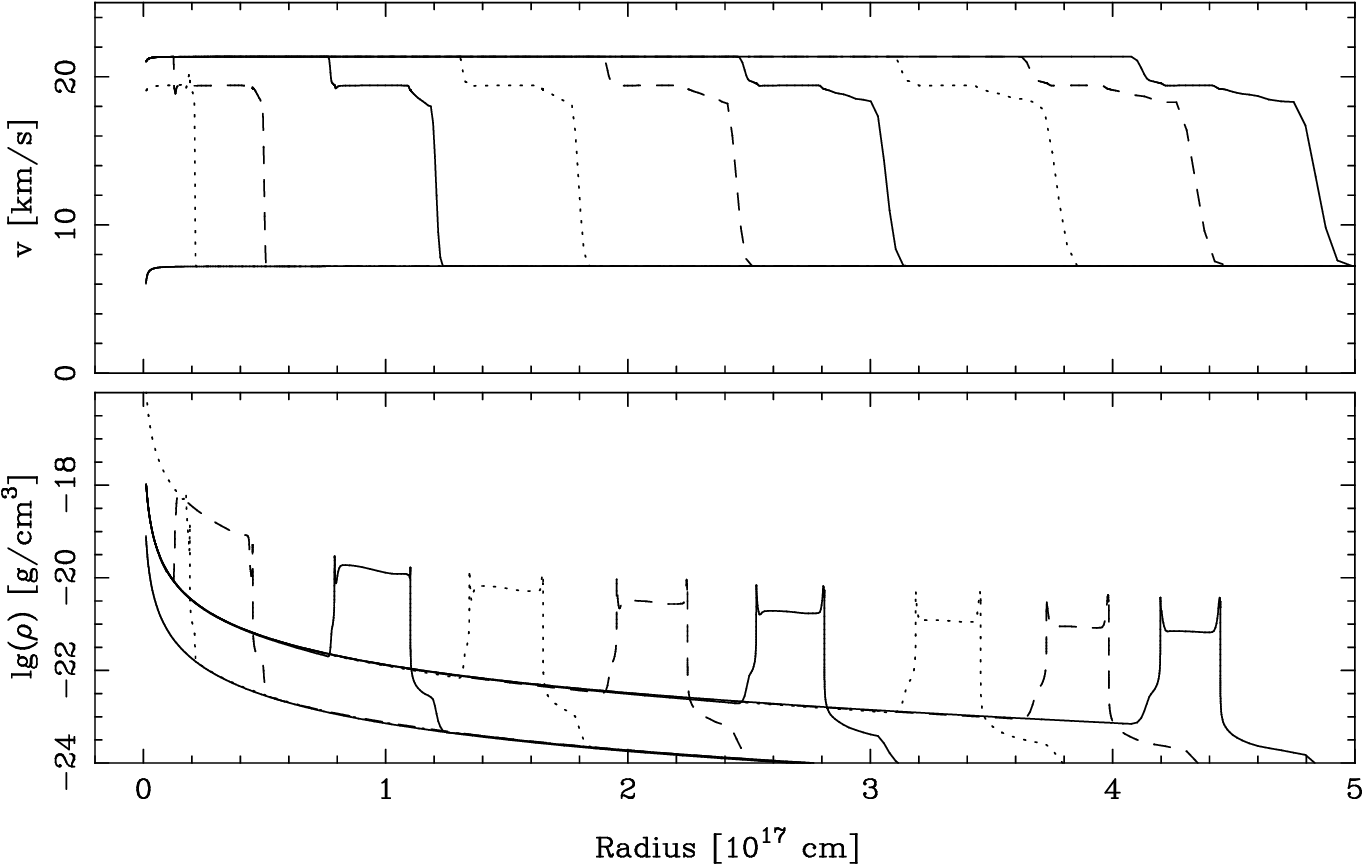}}
  \caption{  \label{testcaseC}
  Nine instants in the evolution of the CSE.
  Upper panel: velocity as function of radius for Case I with the same boundary condition as "case C" in SS2000.
  Lower panel: density as function of radius for Case I with the same boundary condition as "case C" in SS2000.
  }
  \end{figure}

\begin{figure}
  \resizebox{\hsize}{!}{\includegraphics[width=5.7cm, angle=270]{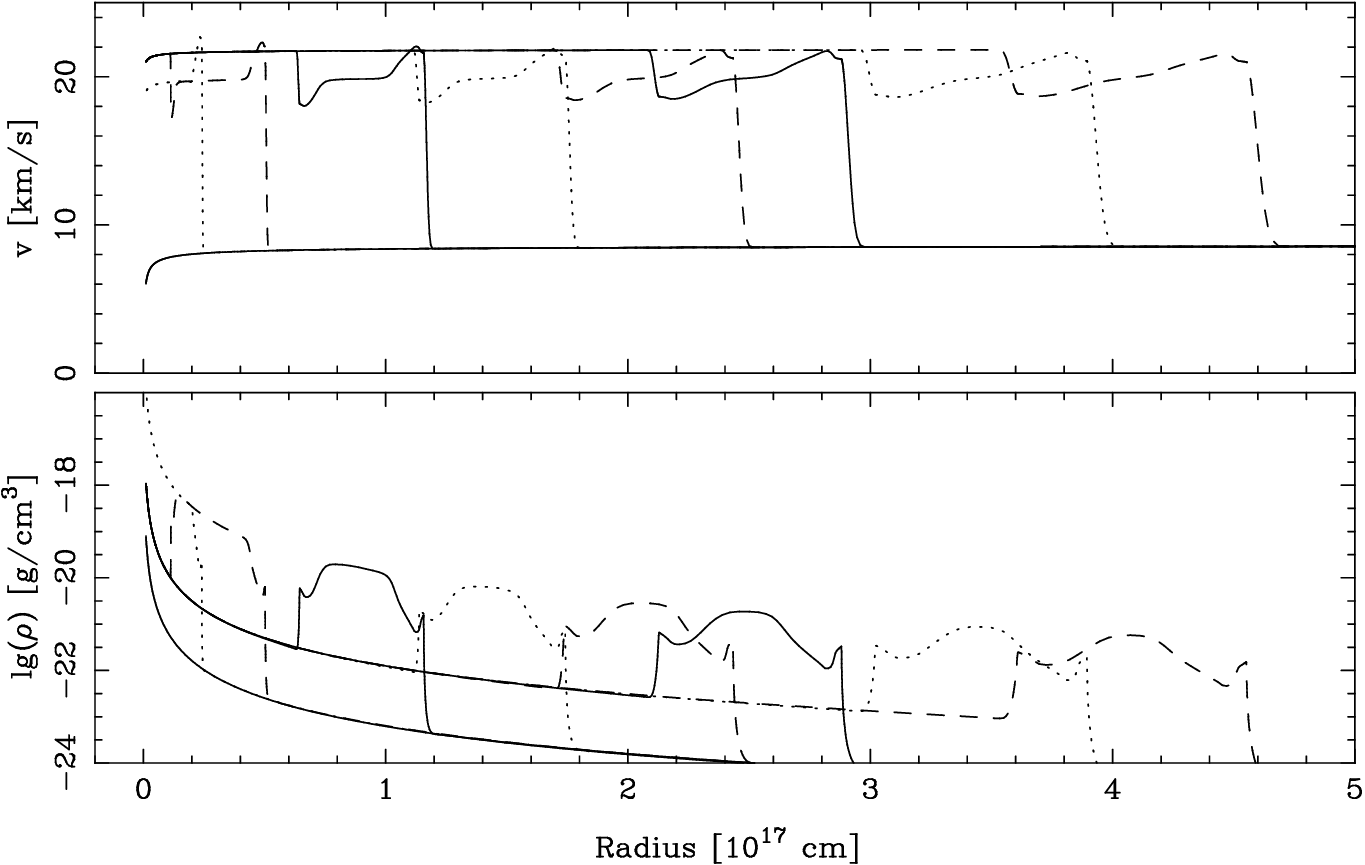}}
  \caption{  \label{testcaseC2}
  Same as Fig. \ref{testcaseC}, but for Case II.}  
  \end{figure}

  Using a simple mass loss evolution of the kind used in SS2000 (i.e. step functions, as in Section \ref{cseres}) in our 
  Case II, which is quite similar to the "NEBEL" models of SS2000, we are able to basically reproduce their results, i.e. 
  the wind velocity field and the mass density of the CSE (Fig. \ref{testcaseB2} and Fig \ref{testcaseC2}) evolve in ways 
  that are very similar to their "case B" and "case C". In Case I we obtain a result that shows some resemblance to the 
  result from the "DEXCEL" model of SS2000, but we see no obvious signs of numerical diffusion. The corresponding plots 
  for Case I (Fig. \ref{testcaseB} and Fig \ref{testcaseC}) are shown for comparison.

\end{appendix}

\end{document}